\documentclass{aa}  
\usepackage{siunitx}
\usepackage{wasysym}
\usepackage{makecell}
\usepackage{adjustbox}
\usepackage{booktabs}
\usepackage{natbib}
\usepackage{float}
\bibpunct{(}{)}{;}{a}{}{,}
\usepackage{longtable}
\usepackage{graphicx}
\usepackage{txfonts}
\usepackage[colorlinks=true, linkcolor=blue, citecolor=blue, urlcolor=blue]{hyperref}

\date{Accepted 2 August 2025}

\usepackage{enumitem}
\usepackage[table]{xcolor}
\DeclareSIUnit{\ME}{M_{\oplus}}
\DeclareSIUnit{\RE}{R_{\oplus}}
\DeclareSIUnit{\RM}{R_{M}}
\DeclareSIUnit{\MM}{M_{M}}
\DeclareSIUnit{\MV}{M_{V}}
\DeclareSIUnit{\Myr}{Myrs}
\DeclareSIUnit{\day}{days}
\DeclareSIUnit{\LEM}{L_{em}}

\begin{document}

 \title{The possibility of a giant impact on Venus}
   \author{M. Bussmann \inst{1} \fnmsep\thanks{Corresponding author: mirco.bussmann@uzh.ch} \and  
          C. Reinhardt \inst{1, 2} \and
          C. Gillmann \inst{3} \and
          T. Meier \inst{1} \and
          J. Stadel \inst{1}\and 
          P. Tackley \inst{3} \and
          R. Helled \inst{1} 
          }

   \institute{Department of Astrophysics, University of Zürich, Winterthurerstrasse 190, CH-8057 Zürich, Switzerland \and
            Space Research \& Planetary Sciences, University of Bern, Sidlerstr. 5, CH-3012 Bern, Switzerland \and
            Department of Earth and Planetary Sciences, ETH Zürich, Sonneggstrasse 5, CH-8092 Zürich, Switzerland 
            }

  \abstract 
    {Giant impacts were common in the early evolution of the Solar System, and it is possible that Venus also experienced an impact.
    A giant impact on Venus could have affected its rotation rate and possibly its thermal evolution.
    In this work, we explored a range of possible impacts using smoothed particle hydrodynamics (SPH). 
    We considered the final major collision, assuming that differentiation already occurred and that Venus consists of an iron core (30\% of Venus' mass) and a forsterite mantle (70\% of Venus' mass).
    We used differentiated impactors with masses ranging from \SI{0.01}{} to \SI{0.1}{\ME}, impact velocities between \SI{10}{} and \SI{15}{\kilo\meter\per\second}, various impact geometries (head-on and oblique), different primordial thermal profiles, and a range of pre-impact rotation rates of Venus.
    We analysed the post-impact rotation periods and debris disc masses to identify scenarios that can reproduce Venus' present-day characteristics.
    Our findings show that a wide range of impact scenarios are consistent with Venus' current rotation. 
    These include head-on collisions on a non-rotating Venus and oblique, hit-and-run impacts by Mars-sized bodies on a rotating Venus.
    Importantly, collisions that match Venus' present-day rotation rate typically produce minimal debris discs residing within Venus' synchronous orbit.
    This suggests that the material would likely reaccrete onto the planet, preventing the formation of long-lasting satellites — which is consistent with Venus' lack of a moon.
    We conclude that a giant impact can be consistent with both Venus' unusual rotation and lack of a moon, potentially setting the stage for its subsequent thermal evolution.
    }

    \maketitle

\section{Introduction}

Among the planets in the Solar System, Venus is the most similar to Earth in mass, radius, bulk density, and semi-major axis. 
Despite these similarities, surface conditions on Venus are vastly different from those on Earth, with surface pressures over 90 times that of Earth, average surface temperatures of \SI{740}{\kelvin}, an atmosphere composed mainly of carbon dioxide, and clouds of sulfuric acid. In addition, Venus rotates very slowly in a retrograde direction and has no moon. 
\\
It remains unknown why Venus and Earth are so different from each other. Answering this question is important for understanding the diversity of terrestrial planets and planetary habitability. 
It is also unclear whether the two planets were already distinct after their formation or started with similar conditions and diverged during their evolution.
Many different pathways for Venus' evolution have been suggested \citep{gillmannLongtermEvolutionAtmosphere2022} and range from a long-lived primordial magma ocean phase without habitability \citep{hamanoEmergenceTwoTypes2013} to paths that describe Venus as habitable for most of its lifetime with a surface water ocean, plate tectonics, and stable temperate conditions \citep{wayVenusianHabitableClimate2020}.
A key measurement in that aspect is the high deuterium-to-hydrogen ratio in Venus' atmosphere \citep{donahueVenusWasWet1982}, which is currently the highest measured in the Solar System. 
This may indicate that Venus experienced extensive hydrogen escape and thus likely had a larger water inventory \citep{raymondHighresolutionSimulationsFinal2006,  gillmannConsistentPictureEarly2009}.
However, Venus' early state and its evolution are still poorly constrained, and more data and further investigations are required.
\\

Generally, the evolution and formation of terrestrial planets is determined and affected by various processes.
Among them, giant impacts can change the planetary structure and bulk composition, as well as planetary rotation and the presence of moons.
Indeed, many of the observed properties of the planets in the Solar System are thought to be a result of giant impacts. 
These include Mercury’s high iron-to-rock ratio (e.g. \citealt{benzCollisionalStrippingMercurys1988,  asphaug_mercury_2014, chauFormingMercuryGiant2018}), Mars' crustal dichotomy (e.g. \citealt{wilhelmsMartianHemisphericDichotomy1984, chengCombinedImpactInterior2024}), and the formation of Earth's Moon (e.g. \citealt{hartmannSatellitesizedPlanetesimalsLunar1975, cameronOriginMoon1976}). 
In addition, dynamical simulations of the early evolution of the Solar System clearly show that giant impacts were very common \citep[e.g.][]{asphaug_mercury_2014}. 
In the case of Venus, no obvious feature indicates that it suffered a giant impact, but statistically speaking it is rather likely that such an impact did occur \citep{chambersMakingMoreTerrestrial2001, quintanaFrequencyGiantImpacts2016}.
It is therefore interesting to explore what type of impacts could be consistent with Venus' physical properties, in particular, its rotation period and lack of a moon.
\\

Giant impacts can change the angular momentum of the planet and also lead to the formation of a moon.
They have been suggested to play a crucial role in the deceleration of Venus' spin rate towards the present-day state. 
The present-day rotation period of Venus is consistent with tidal deceleration from gravitational and thermal atmospheric tides \citep{correiaFourFinalRotation2001, revolSpinEvolutionVenuslike2023}.
Since this is a very slow process, only sufficiently high initial rotation periods would lead to Venus' present rotation rate in adequate time.
Depending on the tidal dissipation model, the minimal allowed initial rotation periods range from \SI{.5}{} to \SI{3}{\day} \citep{correiaLongtermEvolutionSpin2003, correiaLongtermEvolutionSpin2003a,musseauViscosityVenusMantle2024}; although, initial rotation periods shorter than two days require specific conditions such as the presence of a long-lasting global water ocean.
Thus, as a conservative constraint, we consider that a giant impact is consistent with Venus' present rotation rate if the post-impact rotation period is two days or longer. Whether a giant impact can lead to such a post-impact spin state critically depends on the rotation of Venus before the collision, which is poorly constrained and depends on Venus' accretion history \citep[e.g.][]{lissauerOriginSystematicComponent1991,morbidelliDidTerrestrialPlanets2024}. 
\\

The formation of a large moon via a giant impact primarily requires the generation of a substantial circumplanetary debris disc with sufficient mass. Typically, oblique collisions involving relatively massive bodies at low velocities provide the necessary mass and angular momentum to generate such discs \citep{timpeSystematicSurveyMoonforming2023, meierSystematicSurveyMoonforming2025}. If the debris then remains in orbit and is not reaccreted or lost due to dynamical or tidal effects, it can coalesce into a moon. The fact that Venus, unlike Earth, has no large moon implies that if the planet suffered from a giant impact it occurred in conditions that disfavour the formation of a large moon (e.g. small debris disc or significant material loss). 
To what extent a smaller impact consistent with Venus' very low spin angular momentum could generate a small debris disc that, via viscous spreading, resulted in the formation of one or several small moons \citep{charnoz_recent_2010, hyodo_formation_2015} has not been investigated yet. However, these moons are likely to be within Venus' synchronous orbit and therefore fall back onto the planet as the Martian moons  \citep{rosenblattAccretionPhobosDeimos2016}.

In this work, we investigated whether (and under what impact conditions) a giant impact is consistent with Venus' rotation rate and lack of a moon.
A follow-up paper will focus on Venus' subsequent long-term evolution.
Here, we lay the groundwork by establishing the initial thermal state to be used in evolution simulations. 
The paper is structured as follows. 
In Section~\ref{sec:methods}, we describe the methods used to build the models, run the giant impact simulations, and process their results. 
In Section~\ref{sec:resultsanddiscussion}, we present and discuss the results of our simulations. 
Finally, in Section~\ref{sec:summaryandconclusion}, we summarise our findings and conclude whether giant impacts are consistent with Venus' current state.

\section{Methods}
\label{sec:methods}

In this paper, we present a large suite of 3D impact simulations using the smoothed particle hydrodynamics (SPH) method \citep{monaghanSmoothedParticleHydrodynamics1992}. 
The SPH method exhibits excellent conservation properties, can handle geometries with high levels of deformation, and tracks the fate of the material due to its Lagrangian nature, making it a common choice for modelling impacts \citep{canupOriginMoonGiant2001, hyodoImpactOriginPhobos2017, kegerreis_consequences_2018, reinhardtBifurcationHistoryUranus2020, woo_did_2022, timpeSystematicSurveyMoonforming2023, Dou2024, ballantyneInvestigatingFeasibilityImpactinduced2023, meierSystematicSurveyMoonforming2025, Meier2025, denton2025}. 
The simulations were performed using a novel high-performance computing implementation (Meier et al., in prep) of SPH within the gravity code \texttt{pkdgrav3} \citep{potterPKDGRAV3TrillionParticle2017}, derived from the Lagrangian \citep{springelCosmologicalSmoothedParticle2002,priceSmoothedParticleHydrodynamics2012}, that accounts for corrections due to a variable smoothing length and contains kernels that do not suffer from the pairing instability and allow accurate sampling of the fluid \citep{dehnenImprovingConvergenceSmoothed2012}. The code includes state-of-the-art improvements for modelling giant impacts such as an interface and free-surface correction \citep{reinhardtNumericalAspectsGiant2017,reinhardtBifurcationHistoryUranus2020,ruiz-bonillaDealingDensityDiscontinuities2022}, strict conservation of entropy in adiabatic flows \citep{reinhardtNumericalAspectsGiant2017} and a generalised EOS interface \citep{meierEOSResolutionConspiracy2021,meierEOSlib2021}, which simplifies the use of different equations of state and allows the use of multiple EOSs in the same simulation.

\subsection{Models}
\label{sec:models}
\begin{figure}
   \centering
   \includegraphics[width=\hsize]{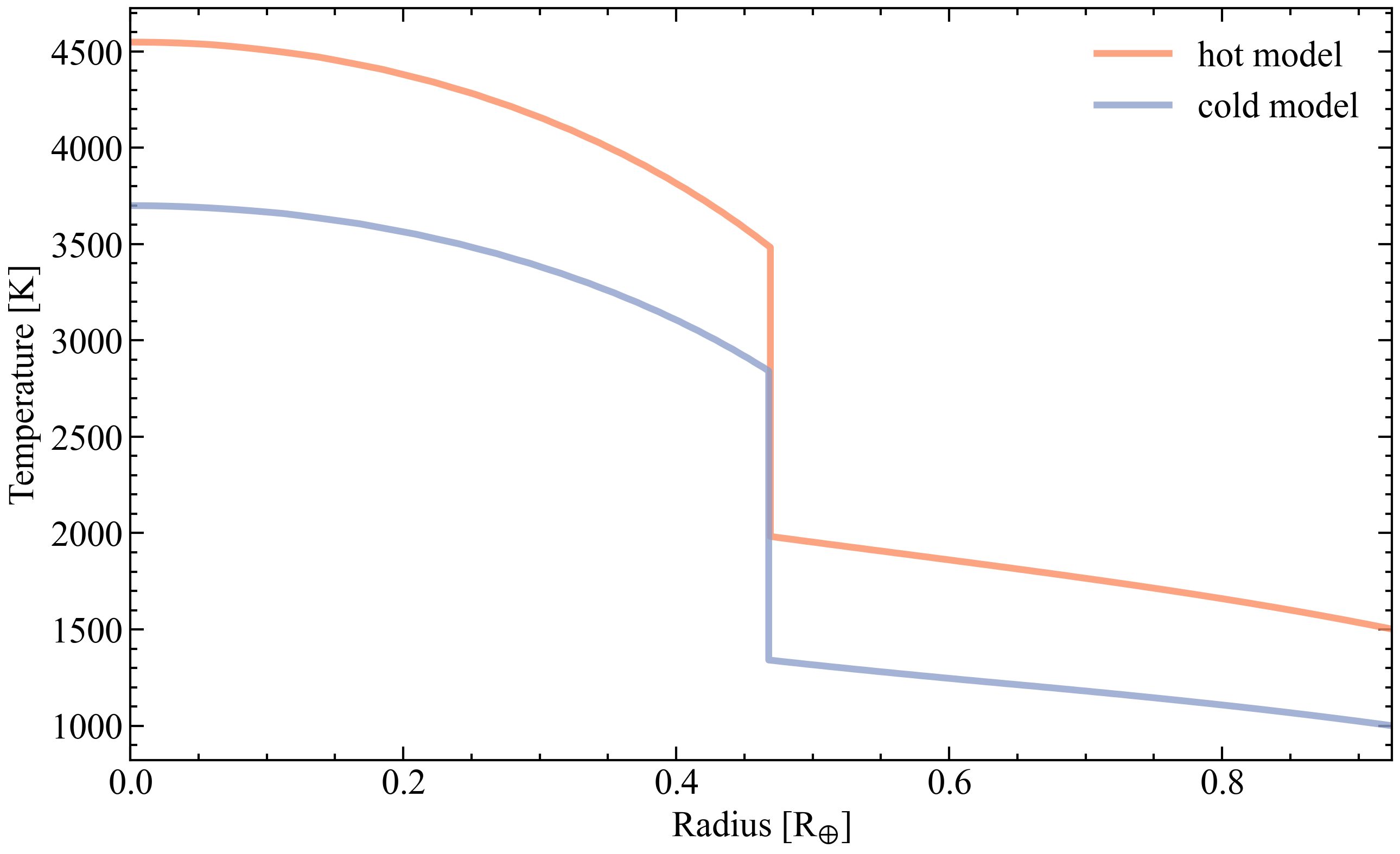}
      \caption{Pre-impact temperature profiles of Venus. The blue and orange curves correspond to the cold model and hot model with surface temperatures of \SI{1000}{\kelvin} and \SI{1500}{\kelvin}, respectively. Both models assume a differentiated planet with an iron core (30\% by mass) and a  rocky mantle (70\% by mass) and include a temperature jump of \SI{1500}{\kelvin} at the core-mantle boundary.
    }
    \label{fig:models}
\end{figure}

We generated 1D adiabatic profiles with iron fraction, surface temperature, and mass as input parameters (Fig.~\ref{fig:models}). 
All Venus models are assumed to be differentiated and consist of an iron core and a rocky mantle, with 30\% and 70\% of the total planetary mass, respectively. 
Since the surface temperature of young Venus is poorly constrained and also depends on the timing of the impact, we considered two different Venus models. The first has a surface temperature of \SI{1000}{\kelvin} (referred to as the cold model) and the second has one of \SI{1500}{\kelvin} (referred to as the hot model). Since a purely adiabatic thermal profile leads to unrealistically low core temperatures, we included a temperature jump at the core-mantle boundary of \SI{1500}{\kelvin} \citep{lay2008cmb}.
We considered a range of pre-impact masses for Venus from \SI{0.705}{} to \SI{0.815}{\ME}. The pre-impact mass is adjusted for each collision so that the post-impact mass is consistent with Venus' present mass within \SI{2.5}{\percent}. We assume that the impactors are differentiated and have the same bulk composition as Venus with a surface temperature of \SI{1000}{\kelvin}. Iron was modelled with the equation of state (EOS) from \citet{stewartEquationStateModel2020}, rock with the forsterite EOS from \citet{stewartEquationStateModel2019}, both of which are constructed from the latest version of M-ANEOS \citep{thompsonImprovementsChartRadiationhydrodynamic1972, meloshHydrocodeEquationState2007, thompsonMANEOS2019}.
\\

We used \texttt{ballic} \citep{reinhardtNumericalAspectsGiant2017} to build low-noise particle representations of the equilibrium models. Each Venus model is resolved with $8 \times 10^5$ particles, while the impactor resolution is adjusted to achieve equal-mass particles. Such a resolution allows us to faithfully predict the post-impact mass of Venus \citep{meierEOSResolutionConspiracy2021} and, if present, the circumplanetary disc \citep{hosono_unconvergence_2017}.
All models were relaxed for a simulation time of \SI{17}{\hour} to further reduce noise and ensure stability. Since the rotation of Venus prior to a late giant impact is not known, we account for different pre-impact spin periods, $T$, of \SI{0}{}, \SI{2.5}{}, \SI{6}{}, \SI{12}{} and \SI{24}{\hour}, whereas a spin period of \SI{0}{\hour} corresponds to no rotation. The fast-spinning case of \SI{2.5}{\hour} translates to about 1.7 $T_{crit}$, where $T_{crit}$ is the critical period for rotational breakup. For a given rotation period, the rotating bodies are generated as described in \citet{meierSystematicSurveyMoonforming2025}.

\subsection{Initial conditions}
\label{sec:initial_conditions}
The collisions are described by the impactor mass, the impact velocity, the impact parameter, and the pre-impact thermal profile of Venus (Table \ref{tab:collisions}). We included impactor masses of \SI{0.1}{} and \SI{0.01}{\ME}. This corresponds to an expected range of impactor masses that collide with a Venus-like body in N-body simulations \citep[e.g.][]{emsenhuberCollisionChainsTerrestrial2023}.
We consider impact velocities of \SI{10}{} and \SI{15}{\kilo\meter\per\second}, which correspond to $\sim 1-1.5$ times Venus' escape velocity. 
The impact parameter describes the geometrical setup of the collision. It was assumed that the collision always occurs on the equatorial plane. Therefore, only one parameter is needed to fully describe the geometric setup of the collision. The impact parameter is defined as the sine of the impact angle. A head-on collision with an impact angle of zero therefore corresponds to an impact parameter of zero. An impact parameter of unity only leads to a tidal interaction between the two bodies. 
We included impact parameters of $0$, $\pm 0.3$, and $\pm 0.7$, which corresponds to impact angles of about 0, 20, and \SI{45}{\degree}.
If Venus is rotating prior to the collision, the orbital and spin angular momentum can be either parallel or anti-parallel. A positive (or negative) impact parameter corresponds to an orbital angular momentum that is parallel (or anti-parallel) to Venus' spin and leads to an increase (or decrease) in the bound angular momentum, respectively.

\subsection{Post-impact Venus}

In order to determine the number of remaining bodies and Venus' mass, following prior work (e.g. \citealt{chauFormingMercuryGiant2018, reinhardtSimulatingGiantImpacts2020, reinhardt_forming_2022,
timpeSystematicSurveyMoonforming2023, meierSystematicSurveyMoonforming2025}), we used the group finder \texttt{skid} \citep{stadelCosmologicalNbodySimulations2001}.
\texttt{Skid} identifies coherent, gravitationally bound clumps of material by identifying regions that are bound by a critical surface in the density gradient and then removes the least bound particles one by one from the resulting structure until all particles are self-bound.
The relevant input parameters are \texttt{nSmooth} which specifies the number of neighbours used to calculate the density and the linking length \texttt{tau} used to determine friends of friends.
We used \texttt{nSmooth} = 3200 and \texttt{tau} = 0.06 and find that the inferred bound mass is insensitive to the choice of \texttt{tau} and the number of smoothing neighbours as long as \texttt{nSmooth} is $ > 1000$. Lower values of \texttt{nSmooth} can result in spurious groups due to noise in the density gradient.
We then categorised the collisions into three outcome scenarios: merger, graze-and-merge, and hit-and-run.
Mergers are collisions in which at least \SI{95}{\percent} of the total colliding mass remains gravitationally bound to Venus post-impact.
A special type of merging collision is the graze-and-merge. In this case, the impactor grazes Venus in the first collision, remains gravitationally bound, and finally collides with Venus again and merges. In a hit-and-run scenario, the impactor skims Venus but does not stay gravitationally bound to Venus afterwards. Thus, there are two intact bodies after the collision that are not gravitationally bound to each other.
\\

The post-impact rotation period of Venus is determined by the following procedure (see \citealt{reinhardtBifurcationHistoryUranus2020} for further details). First, we determined the specific angular momentum and radial positions for all SPH particles that belong to Venus. 
Then, we determined Venus' rotational angular velocity by fitting a uniform rotation profile with $l = \omega r^2$ to the particle data. From this fit, we determined Venus' period using $P = \omega / 2\pi$. 
After a collision, Venus is expected to be very hot and extended. As it cools and contracts over time, its moment of inertia also decreases over time. 
Since angular momentum is conserved, the rotation periods are expected to decrease, and the values inferred from our simulations serve as an upper limit.
\\

For each collision, we inferred the mass of the circumplanetary disc using an iterative algorithm following \citet{canupScalingRelationshipSatelliteForming2001}.
For each step, the planet's radius is determined from the planetary mass and assumed mean density, which is an input parameter. All particles are then either bound or escaping. Bound particles are either assigned to the planet if their periapsis is smaller than the planetary radius or to the disc. This procedure is repeated until the planet, disc, and escaping mass have converged. As input parameters, we used Venus' mean density and a moment of inertia of 0.337 \citep{margotSpinStateMoment2021}.

\section{Results and discussion}
\label{sec:resultsanddiscussion}
We performed a total of 81 simulations and find 64 merger, seven graze-and-merge, and ten hit-and-run simulations. All collisions are listed in Table \ref{tab:collisions}. In the following, collisions are referenced by their simulation id, which is the first column of the table. 

\subsection{Rotation rates}
\label{sec:rotationrates}
\begin{figure}
    \resizebox{\hsize}{!}{%
        \includegraphics{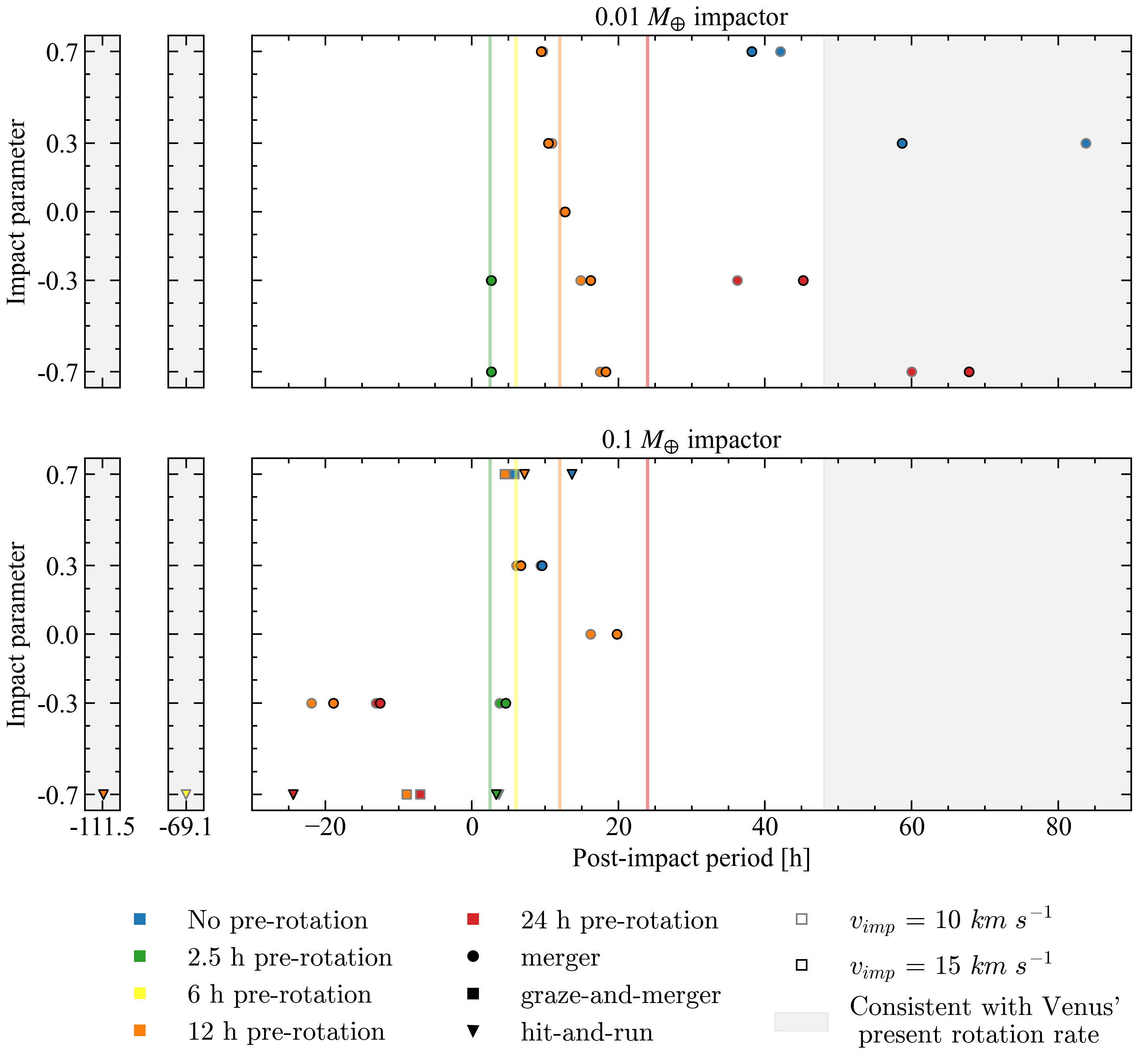}
    }
    \caption{Post-impact rotation period of Venus of the 'cold models' inferred from our simulations. The upper panel shows collisions with a \SI{0.01}{\ME} impactor, while the lower panel shows collisions with a \SI{0.1}{\ME} impactor. The marker colour describes the pre-impact rotation period, the edge colour the impact velocity, and the marker symbol the collision outcome according to the legend. The different pre-impact rotation periods are also displayed with the respective coloured vertical lines. The grey area corresponds to periods that are consistent with Venus' present-day period.}
    \label{fig:periods}
\end{figure}

\subsubsection{Angular momentum transport via collisions}
Figure~\ref{fig:periods} shows the post-impact periods inferred from our simulations for the 'cold' Venus models. The results for the 'hot' models are not shown in the figure because they are almost identical. The head-on impacts on a non-rotating Venus are excluded because the resulting rotation is zero, and non-zero values are numerical artefacts. As expected, we find that collisions of massive impactors with higher impact parameters and velocities (which have higher orbital angular momentum) tend to transfer more angular momentum to Venus.
However, as the orbital angular momentum increases, the collision outcome transits from a merging (and graze-and-merge) to hit-and-run scenario, and not all mass and angular momentum is accreted by Venus.
Graze-and-merge collisions, which lie in the transitional regime between merger and hit-and-run events, represent the scenarios with the highest angular momentum that still result in the complete accretion of mass and angular momentum by Venus.
\\

In our set of simulations, all collisions involving a \SI{0.01}{\ME} impactor lead to mergers.
For the more massive \SI{0.1}{\ME} impactor, the outcome of the collision depends on the impact angle and velocity. Low-impact parameters with $b \leq 0.3$ eject no more than 3\% of the total mass and are therefore also classified as mergers (see Section~\ref{sec:methods} for details).
For an impact parameter of 0.7, we observe that collisions with velocities of \SI{10}{\kilo\meter\per\second} result in graze-and-merge collisions, while higher velocities of \SI{15}{\kilo\meter\per\second} lead to hit-and-run events.
As a result, the highest angular-momentum transfer occurs for \SI{10}{\kilo\meter\per\second} (i.e. simulation 3), and not for \SI{15}{\kilo\meter\per\second}. If Venus did not rotate prior to the giant impact, this collision would lead to a rotation period of \SI{5.8}{\hour}. 
This trend deviates for Venus with pre-impact rotation periods of \SI{2.5}{} and \SI{6}{\hour}, where the collisions with an impact velocity of \SI{10}{\kilo\meter\per\second} also lead to hit-and-run scenarios. This is likely due to flattening and extension caused by the fast rotation, which changes Venus' shape and therefore its cross-section. We find that the amount of angular momentum transferred is quite sensitive to the impact velocity and the impact parameter. The temperature, on the other hand, has little effect on the post-impact rotation rate.

\subsubsection{The importance of pre-impact rotation}
When an impactor collides with an already rotating Venus, it may modify Venus' angular momentum and rotation period. 
We find that the amount of angular momentum transferred during a giant impact is rather insensitive to the pre-impact rotation and that identical collisions on Venus with different rotation periods result in the same angular momentum transferred. 
Only for two collisions with initial rotation periods of \SI{2.5}{} and \SI{6}{\hour}, it deviates from this behaviour due to the different outcome scenario, as discussed above. 
Depending on how the orbital and spin angular momentum are aligned, a collision can increase or decrease the planet's rotation period. 
A positive impact parameter increases the spin angular momentum and Venus' rotation, while a negative impact parameter has the opposite effect. 
While angular momentum remains constant in head-on collisions ($b=0$), the spin periods may still change due to modifications in Venus' moment of inertia as a result of accretion of mass and shock heating. 
We determined the post-impact rotation periods only days after the giant impact when Venus is still very hot and extended, and we expect it to cool and contract over longer timescales than those covered in the simulations, which decreases the planetary rotation period. 
However, as Venus' mass has increased significantly, even after cooling the equilibrated periods are expected to be different from the pre-impact values. 
As an example, simulation 12, which has a pre-impact rotation period of \SI{12}{\hour}, results in a post-impact period of \SI{19.8}{\hour}, but the final rotation period after cooling is expected to be $\sim \SI{15}{\hour}$.
To verify that the qualitative results remain consistent, we also examined our conclusions with regard to the post-impact angular momentum (see Fig.~\ref{fig:AM_plot} in Appendix), and it is indeed in agreement.

\subsubsection{Impacts on Venus without pre-impact rotation}
\begin{figure*}
   \centering
   \includegraphics[width=\hsize]{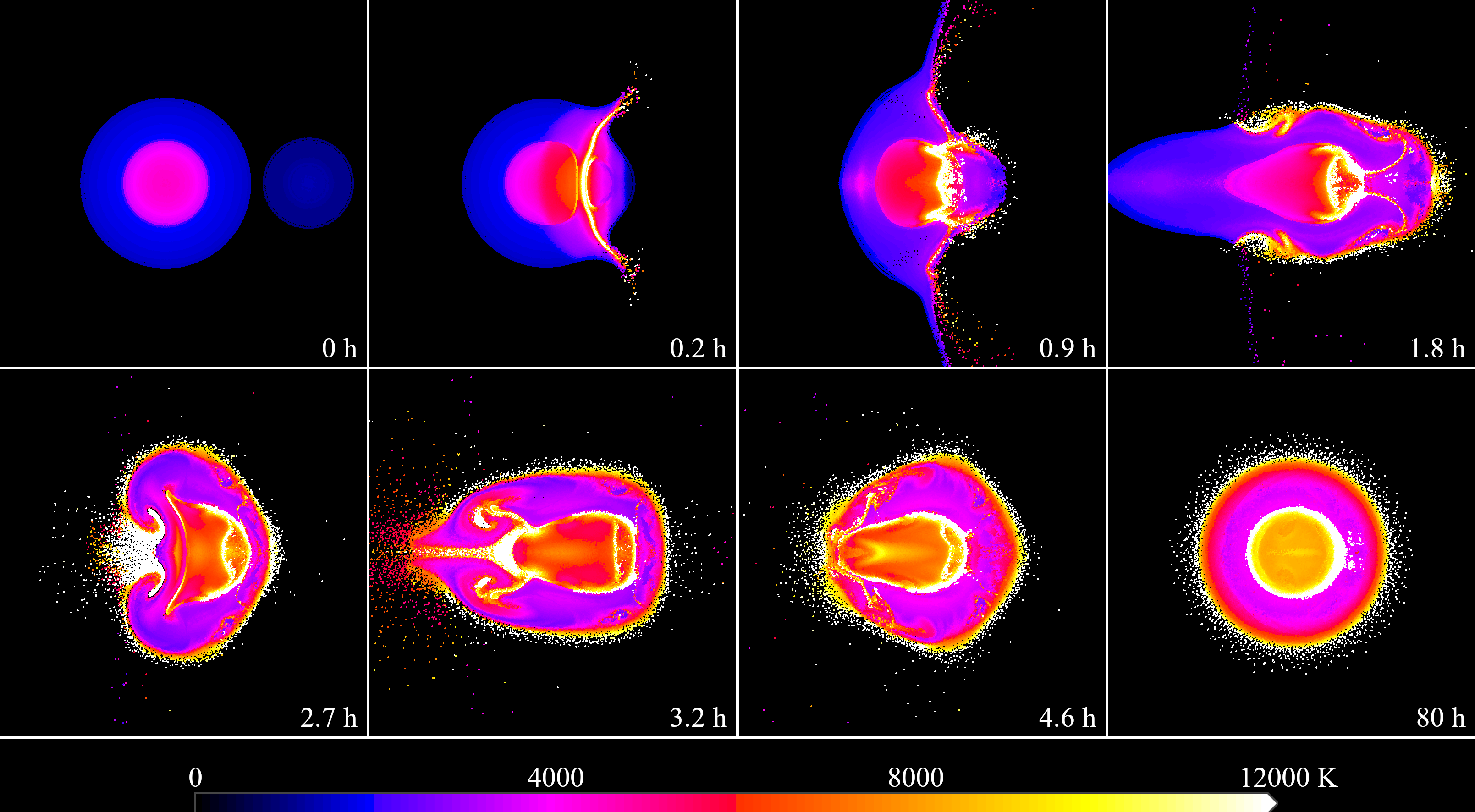}
      \caption{Snapshots of a cross-sectional slice of a head-on collision between a non-rotating Venus and a \SI{0.1}{\ME} impactor at \SI{10}{\kilo\meter\per\second}, shown at multiple time steps. Initial energy deposition at the impact site generates pressure waves that converge at the antipode, causing significant heating and deformation. The outcome of this collisions is a merger.}
    \label{fig:snapshots_headon}
\end{figure*}

For a giant impact to be consistent with Venus' current rotation rate, the post-impact rotation period must be greater than \SI{48}{\hour}. 
We find 17 collisions that are consistent with this constraint (green shaded collisions in Tab. \ref{tab:collisions}). 
Without pre-impact rotation, the giant impact is required to transfer very little to no angular momentum, a condition that is evidently satisfied by head-on collisions.
Fig.~\ref{fig:snapshots_headon} shows the cross-sectional temperature profile of such a collision with a \SI{0.1}{\ME} impactor and an impact velocity of \SI{10}{\kilo\meter\per\second} (i.e. simulation 1).
Additionally, for small impactors with \SI{0.01}{\ME}, collisions with low-impact parameters can also be compatible with Venus' present rotation. 
Such collisions include those with an impact parameter of 0.3 and impact velocities of \SI{10}{} and \SI{15}{\kilo\meter\per\second} (simulations 43 and 46), resulting in post-impact rotation periods of \SI{84}{} and \SI{59}{\hour}, respectively. 
In contrast, with an impact parameter of 0.7 (e.g. simulation 44), the angular momentum contribution exceeds the threshold, leading to rotation periods of \SI{42}{\hour} or less. 
For large impactors (\SI{0.1}{\ME}), only head-on collisions are consistent with Venus' rotation. 
Even a low-velocity collision with an impact parameter of 0.3 (simulation 2) results in a post-impact rotation period of \SI{9.4}{\hour}. 
Interestingly, all consistent collisions without pre-impact rotation lead to mergers and never hit-and-run or graze-and-merge events, because these scenarios deposit too much angular momentum.

\subsubsection{Impacts on Venus with pre-impact rotation}

\begin{figure*}
   \centering
   \includegraphics[width=\hsize]{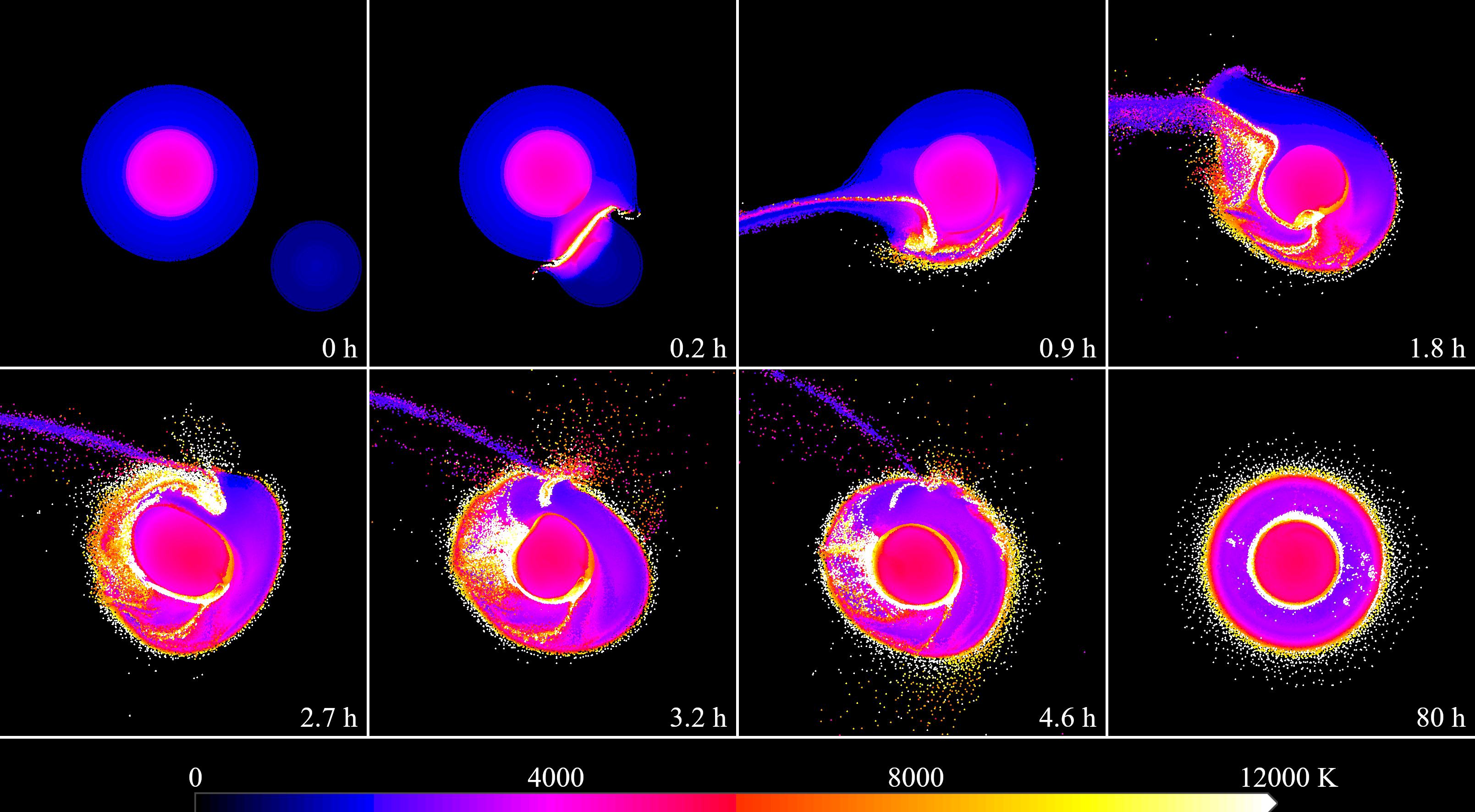}
      \caption{Snapshots of a cross-sectional slice of an oblique collision (impact parameter -0.7) between a rotating Venus (6-hour period) and a \SI{0.1}{\ME} impactor at \SI{10}{\kilo\meter\per\second}, shown at multiple time steps. The impactor grazes the planet and escapes with minimal disruption, classifying this as a hit-and-run event.}
    \label{fig:snapshots_oblique}
\end{figure*}

When considering pre-impact rotation, the transferred angular momentum has to approximately cancel the spin angular momentum to be consistent with Venus' present rotation period. 
This leads to a broader parameter space for successful collisions, as they are not limited to (near) head-on collisions.
This allows us to not only identify merging collisions, but also to observe that some hit-and-run collisions can explain Venus' present rotation rate. 
We find that our \SI{0.01}{\ME} impactor collisions have insufficient orbital angular momentum to slow down a pre-impact rotation period of \SI{12}{\hour} or less. 
In particular, the collision with the highest orbital angular momentum (simulation 57) increased a pre-impact rotation period of \SI{12}{} to only \SI{18}{\hour}, remaining well below \SI{48}{\hour}. 
In contrast, such collisions are consistent with a 48+~$h$ post-impact rotation period if the pre-impact rotation period was \SI{24}{\hour} or longer.
We find that \SI{0.01}{\ME} impactor collisions with an impact parameter of -0.7 and impact velocities of \SI{10}{} and \SI{15}{\kilo\meter\per\second} on Venus with a pre-impact rotation period of \SI{24}{\hour} (simulations 59 and 61) are successful and lead to post-impact rotation periods of \SI{60}{} and \SI{68}{\hour}, respectively.
Collisions with the more massive impactor (\SI{0.1}{\ME}) generally transfer more angular momentum and therefore are able to slow down a faster rotating Venus.
Specifically, we find simulations 16 and 17 slow down Venus with rotation periods of \SI{12}{} and \SI{6}{\hour}.
The post-impact rotation periods after these collisions are \SI{-111}{} and \SI{-69}{\hour}, so they even lead to slightly retrograde motion, very similar to Venus' present rotation.
Snapshots of simulation 17 are shown in Fig.~\ref{fig:snapshots_oblique}.
Finally, we find that no collision was able to slow down a Venus with a pre-impact rotation period of \SI{2.5}{\hour}. 
We ran some additional simulations with higher impact velocities of \SI{20}{} and \SI{25}{\kilo\meter\per\second}, and even these could not even remotely slow down Venus (i.e. with a pre-impact rotation period of \SI{2.5}{\hour}). 
Thus, if Venus rotated near the break-up velocity, it must have experienced either multiple, very energetic giant impacts with an opposing orbital angular momentum, or giant impacts with an impactor-to-target ratio close to one.

\subsection{Circumplanetary disc mass}
\begin{figure}
    \resizebox{\hsize}{!}{%
        \includegraphics{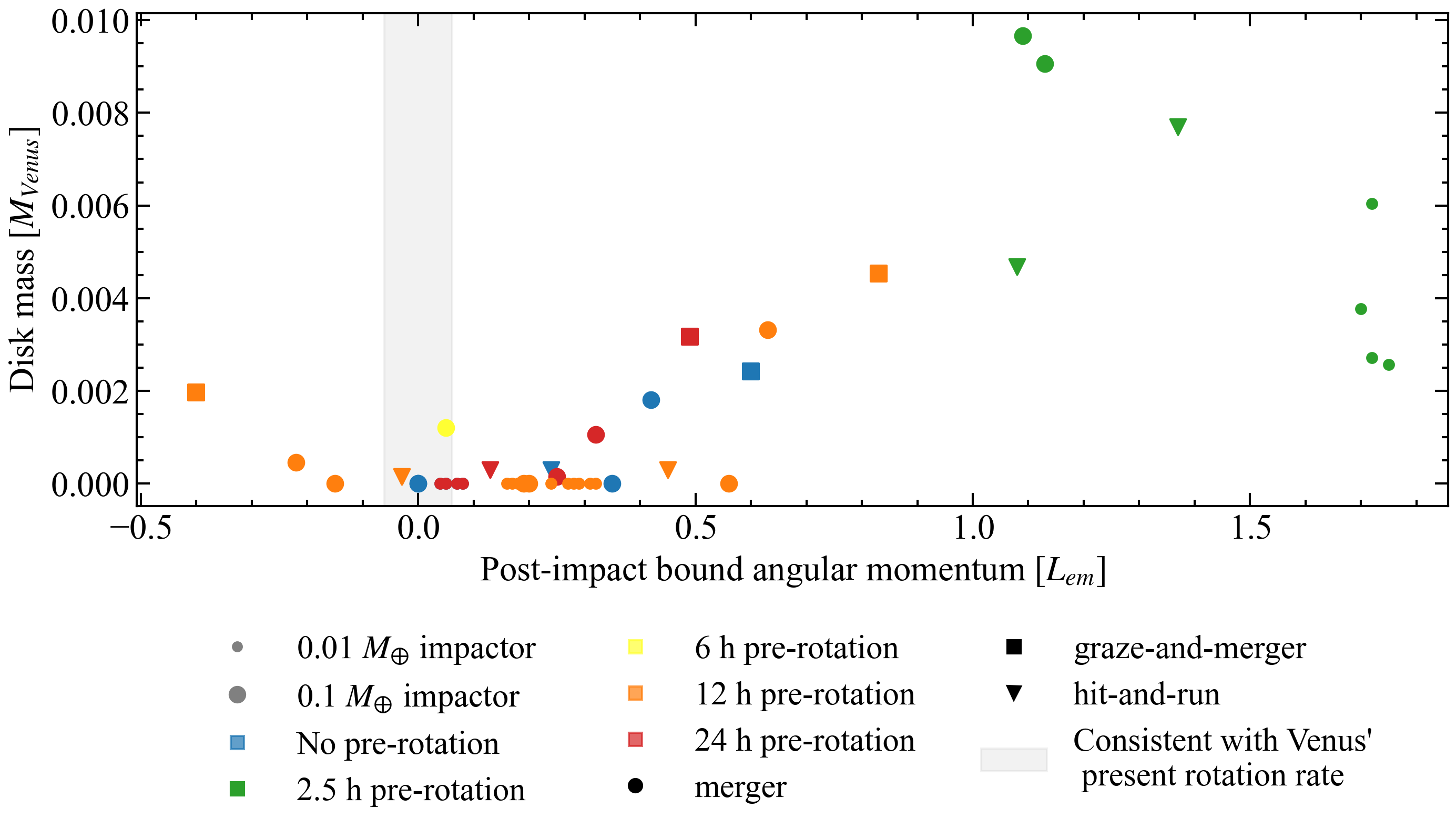}
    }
    \caption{Disc mass generated for different collisions and plotted according to the total post-impact bound angular momentum (planet + disc) in units of the angular momentum of the system Earth-Moon, $L_{em}$. The marker colour describes the pre-impact rotation period and the marker size the impactor mass according to the legend. The marker symbol describes the collision outcome. The greyed area corresponds to the post-impact angular momentum that is consistent with Venus' present rotation period.}
    \label{fig:disk_mass}
\end{figure}

The formation of a significant debris disc around Venus from a giant impact depends on the impact conditions.
For all collisions, we determined the disc mass using a disc finder (see methods in Section~\ref{sec:methods}).
Clearly, since Venus has no moon, we are interested in collisions that do not lead to massive discs. We also note that for the models considered in this study, the inferred disc mass is insensitive to Venus' pre-impact thermal profile.
We find a strong, positive correlation between disc mass and post-impact angular momentum (see Fig.~\ref{fig:disk_mass}), which is consistent with results of previous studies  \citep[e.g.][]{timpeSystematicSurveyMoonforming2023, meierSystematicSurveyMoonforming2025}.
As a result, massive discs typically form for massive impactors and high impact angles and velocities \citep[e.g.][]{meierSystematicSurveyMoonforming2025}. 
Exceptions are simulations with a pre-impact rotation of \SI{2.5}{\hour,} where all collisions, even if they are head-on or with lower impactor masses, result in massive discs.

On the other hand, for all the other pre-impact rotation periods (including non-rotating), we find that head-on collisions and collisions with the smaller impactor (\SI{0.01}{\ME}), do not lead to disc formation, as required in the case of Venus. 
Similarly to the angular momentum transferred, the disc mass produced in a collision also depends on the collision outcome. 
We find that for a given impactor mass, impact velocity, and pre-impact rotation period, graze-and-merge events lead to the highest disc masses, followed by merger and hit-and-run collisions where the impactor remnant can carry away mass and angular momentum. 
\\

Since Venus has no moon, giant impacts that do not produce massive discs are favoured. 
The disc mass required to form a moon is not clearly quantifiable as it depends on many factors such as the mass of the moon or accretion efficiency. 
Therefore, we identify collisions that produce low disc masses as more likely to be consistent with observed Venus than collisions that produce massive discs. 
Since disc mass correlates positively with post-impact angular momentum, collisions that are compatible with Venus' present rotation tend to produce lower disc masses. 
Indeed, we find that all collisions that are consistent with Venus' rotation rate produce discs of at most \SI{1.2e-3}{\MV}, where \SI{}{\MV} is the mass of Venus.
Whether a satellite can be formed from such a disc depends on the radial distribution of the mass and especially on whether the material is inside or outside the synchronous orbit.
Since in our scenarios, Venus is required to have a rotation period of at least \SI{48}{\hour} to be consistent, the synchronous rotation radius, $R_{syn}$, is at \SI{9.8}{\RE} or further out as $R_{syn} \propto T^2$.
Shortly after the collision, most of the disc's material is on highly eccentric orbits.
As the disc particles collide with each other, they tend to settle into approximately circular orbits over time.
Thus, we determined the equivalent circular radius that describes circular orbits with the same angular momentum as the initial eccentric orbit (e.g. \citealt{hyodoImpactOriginPhobos2017, canupOriginPhobosDeimos2018}).

For comparison, the Earth-Moon system formed via an oblique collision with a Mars-sized object \citep[e.g.][]{canupOriginMoonGiant2001}. 
The simulations $22$ and $24$ have very similar impact conditions and, indeed, produce massive discs (up to \SI{0.01}{\MV}). However, these collisions leave Venus with an angular momentum of $~\SI{1}{\LEM}$ (angular momentum of the Earth-Moon system), which is more than a factor of ten higher than requested to explain Venus' observed rotation period.

In the case of Earth, to reconcile the isotopic data of lunar samples with a giant impact, several scenarios of high angular momentum were proposed \citep{canup_forming_2012, cukMakingMoonFastspinning2012, lock_structure_2017}. These result in an excess angular momentum of $\sim \SIrange{1}{2}{\LEM}$ and could be removed by evection resonances or chaotic obliquity tides \citep{cukMakingMoonFastspinning2012, cukTidalEvolutionMoon2016, cukTidalEvolutionEarthMoon2021}. However, these models require very specific conditions to be efficient \citep{rufuTidalEvolutionEvection2020, chenTidalDissipationLunar2016} and cannot reduce the angular momentum below \SI{1}{\LEM}, as requested for Venus.

For all our consistent collisions, the entire disc is inside the synchronous orbit, and thus the material is expected to spiral inward due to tidal forces and eventually collide with Venus.
This prevents the formation of a moon and is consistent with the case of Venus. 
Finally, all collisions on Venus with rotation periods comparable to the break-up period lead to massive discs. 
Thus, in this case, in order to explain the lack of a moon, subsequent moon formation must have been hindered, or a moon must have been lost, possibly due to a subsequent giant impact \citep{alemiWhyVenusHas2006}.

\subsection{Melting profiles and implications for Venus' thermal evolution}

The results discussed above highlight that giant impacts that are consistent with Venus' present rotation period and the lack of a moon are very diverse. Very different combinations of size (mass), angle, and velocity could potentially lead to the conditions observed today due to a trade-off between these properties when it comes to the outcome of the collision. A possible additional method for discriminating between these impact scenarios is to investigate the longer term effects of the collision and the physical state and thermal evolution post-impact. In the following, we present the post-impact temperature field and the expected melting. 
\\

\begin{figure*}
   \centering
   \includegraphics[width=\hsize]{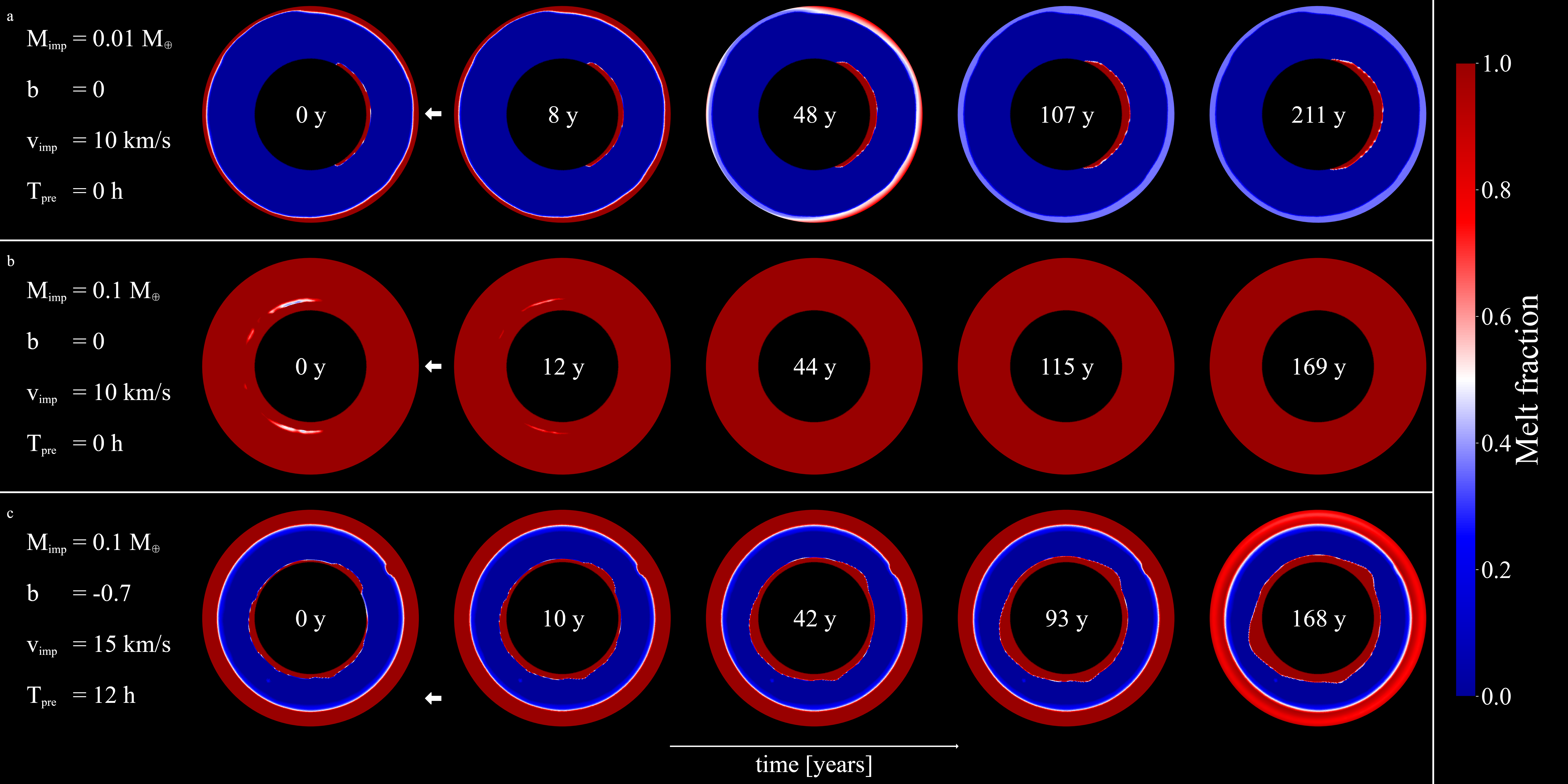}
      \caption{Short-term evolution of Venus' post-impact melting profile for three collisions that can explain its present-day rotation rate and lack of a moon. Impact conditions for each case are listed on the left. The impact locations are shown by white arrows. The colour scale indicates melt fraction, ranging from fully solid (blue) to fully molten (red). Depending on the impact conditions, the resulting melt profiles vary significantly, from a shallow magma ocean confined to the surface to complete mantle melting.}
    \label{fig:melting_profiles}
\end{figure*}

The temperature fields obtained as a result of the SPH giant impact simulations are converted into input files for the mantle dynamics code \texttt{StagYY} (e.g. \citet{tackley2008modelling}), following a similar method to that outlined by \citet{cheng2024combined}. 
First, the data were converted from a 3D particle distribution to a 2D grid representation using \texttt{tipgrid}. Then, the radii of the output grid points, where the local interior temperature is defined, were adjusted to match the normalised mantle mass and radius of a typical \texttt{StagYY} Venus model. The density field is not considered in Fig~\ref{fig:melting_profiles}. \texttt{StagYY} solves for the conservation equations of mass, momentum, and energy for compressible anelastic Stokes flow, utilising an iterative solver. Here, we used a 2D spherical annulus geometry with a resolution of 512 x 96 cells, with radial refinement near the top and bottom boundaries. In the mantle, two mineral phase systems are considered, with their solid-state phase changes (see \citet{tian2023tectonics}): olivine and pyroxene-garnet. Both systems can melt. Rheology was calculated as in \citet{tian2023tectonics} and will be expanded upon in future work when we investigate post-impact mantle dynamics; here, we only focused on the initial state of the mantle immediately after a giant impact and the possible generation of a magma ocean. The solidus and liquidus are depth-dependent. The solidus of the pyrolite (20$\%$ basalt) used in the models is fitted to experimental observations on the anhydrous solidus for peridotite for the upper mantle \citep{herzberg2000new} and high-pressure measurements of the solidus of a glass of pyrolitic composition \citep{mcdonough1995composition} for the lower mantle \citet{zerr1998solidus}. 

Due to the early era considered here and the high temperatures induced by the giant impact, high melt fractions are expected. Silicate melts have very low viscosities, orders of magnitude smaller than those of the solid mantle. The presence of such melts affect the dynamic and thermal evolution, creating volumes of vigorous convection and efficient heat transport and cooling. In regions where the melt fraction exceeds the rheological transition (i.e. solid disaggregates; at 35 \% melt), we used an effective 'eddy' thermal conductivity of 10$ˆ10$ W/(m.K) with a formulation as in \citet{lourencco2020plutonic}, and the heat flux is parametrised following \citet{abe1993, abe1997}. The thermal conductivity is considered to be that of the solid rock when the melt fraction is below the rheological transition (the effective thermal conductivity is then negligible). As a result, for the timescales considered here, that is, immediately after a giant impact and in the following tens to thousands of years, little solid-state convection has time to occur and most of the thermal evolution is caused by diffusion, particularly the effective thermal conductivity treatment representing vigorous magma convection, and melting/freezing. The solidification time refers to the time it takes for a model to reach the rheological transition from an initially liquid state.
\\

Figure~\ref{fig:melting_profiles} shows the resulting 2D annulus melting profile for three impact simulations that are consistent with the expected present-day Venus (1D temperature profiles of them are shown in Fig.~\ref{fig:post_imp_temp_plots}). All three exhibit melting in a surface layer and near the core-mantle boundary because to high temperatures, although the extent of the melting varies considerably between different scenarios: from thin uneven layers (Fig.~\ref{fig:melting_profiles}a), to a fully molten mantle (Fig.~\ref{fig:melting_profiles}b). The merging of the impactor core with the target's core causes the heating of the core and base of the mantle. Re-accretion of hot particles at the surface of the planet results in the surface molten layer. 
At extreme temperatures (from about \SI{6000}{} up to \SI{\sim10000}{\kelvin}), the surface layer is expected to radiate energy efficiently and cool rapidly. Cooling proceeds in three broad phases: i) from a fully molten surface to the rheological transition, ii) from there to pockets of melt and a low melt fraction (a few \%), and iii) onward to a fully solid surface and lithosphere. Fig.~\ref{fig:melting_profiles} shows the first stage (i) of the rapid post-impact cooling. The reader is referred to Appendix D for a brief illustration of stages (ii) and (iii). In cases a and c (see Fig.~\ref{fig:melting_profiles}), the shallow molten layer solidifies efficiently in the first 50 to 200 years, respectively, and the melt fraction at the surface drops below the rheological transition on the order of 2000 yr. A deep magma ocean (case b) stays molten over these timescales and solidifies over a longer time $\approx$ 1000-10000s years for the initial phase \citep{abe1997,monteux2017magmaocean}. In this phase, heat is extracted very quickly from the interior, as expected from the effective 'eddy' thermal conductivity treatment. The observed timescales are consistent with the 1D models of \citet{abe1997} for the deep magma ocean (case b in Fig.~\ref{fig:melting_profiles}).

The deep magma ocean in case b is caused by the high energy transfer and deformation in a head-on collision with an impactor with a larger mass. From a planetary evolution point of view, fully molten mantles resulting from a late giant impact, such as that modelled here, are very similar to the primordial magma oceans generally assumed as precursors to the starting conditions for solid mantle convection numerical models. Therefore, their signatures on long-term evolution, compared to a regular young planet starting with a hot interior, are not yet obvious. 
The core-mantle boundary molten layer can persist for longer and even grow over time since it is heated by the core. However, in the simple conditions tested in Fig.~\ref{fig:melting_profiles}, it becomes unstable, forming buoyant plumes that rise quickly. Over timescales of $\approx$ 1000 years, efficient cooling of the core and base of the mantle occurs. 
\\

A giant impact possibly has implications for the planets' subsequent thermal evolution.
During a collision, the kinetic energy of the impactor is transferred to the target body as mechanical and then thermal energy (see \citealt{okeefe1975}, \citealt{Safronov1978}, \citealt{Kaula1979}, \citealt{Croft1982}, and \citealt{melosh_impact_1989} for the theory).
On Earth-sized target planets, sufficiently energetic impacts cause substantial heating and melting \citep{Jones2002,monteux2007}. 
With large impacts (radius above 100 km), this thermal effect can affect a volume of the planet that can reach the mantle and affect convection patterns \citep{abbott2002,reese2004,watters2009,Roberts2012}. 
Impacts have been proposed as a possible cause for the onset of plate tectonics \citep{maruyama2018}. 
The emplacement of hot buoyant anomalies in the mantle, which break the lithosphere, has been suggested to cause large-scale subduction-like behaviour that rolls back from the impact location and causes downwelling for a limited period of time after a collision \citep{gillmannEffectSingleLarge2016}. 
\citet{oneill2017} have shown that extensive periods of surface mobility could result from successive multiple impacts on a planet. 
\citet{borgeat2022} have shown with 3D simulations that the subduction resulting from series of impacts is extremely important for primitive mantle dynamics, but likely temporary and does not last long after the impact flux dwindled.
\\

Various groups (see above and \citealt{ruedas2019}) have documented the lack of unambiguous long-term signatures of impacts in the thermal evolution of terrestrial planets. 
Giant impacts (radius upwards of 1000 km), however, can release orders of magnitude more energy than the collisions described above, and they would affect a terrestrial planet on a global scale.
Moreover, the geometry of the temperature field resulting from giant impacts is likely different from previously used idealised parametrised treatments and could affect subsequent mantle dynamics. 
In doing so, through mantle convection, they affect the heat transfer and how fast a planet cools down. 
Therefore, the thermal consequences of giant impacts could be critical for their long-term evolution. 
It was recently suggested \citep{marchiLonglivedVolcanicResurfacing2023} that giant impacts could have affected Venus in the long term and caused ongoing volcanism and resurfacing by superheating its core. 
However, it is possible that such an impact could create a global magma ocean and, with large amounts of melt, allow rapid cooling of the interior of the planet and core. 
Investigating the thermal state and subsequent history of Venus post-impact offers a new possibility to constrain the early stages of its evolution, and we plan to investigate this in detail in future research.  
\\

\section{Summary and conclusions}
\label{sec:summaryandconclusion}
We investigated the possibility of giant impacts on Venus and searched for impacts that are consistent with its rotation rate and lack of a moon. 
We considered collisions with different impactor masses, impact velocities, and impact angles.
We used two primordial thermal profiles of Venus and a range of different rotation periods prior to the impact.

We find that a wide range of giant impacts can reproduce  Venus' rotation period. These impacts can have very different   conditions and various outcomes. 
We also find that the generated circumplanetary disc mass correlates positively with the post-impact angular momentum.

Overall, our conclusions can be summarised as follows:
\begin{itemize}
    \item Giant impacts consistent with Venus’ present-day rotation period are very diverse and range from mergers to hit-and-runs, from small to massive impactors and from head-on to oblique collisions.

    \item Such impacts generally produce lower disc masses where most of the material resides inside the synchronous orbit, making long-lived satellites unlikely.

    \item The variety of these collisions leads to very different temperature and melting profiles and thus initial state for its subsequent thermal evolution.

\end{itemize}

The collisions that we investigated are very diverse and would lead to very different thermal and melting profiles of Venus. We plan to explore the implications of such impacts on Venus' subsequent thermal evolution in a follow-up study.
\par 

Overall, our understanding of Venus is expected to improve significantly in the near future thanks to the EnVision (ESA), DAVINCI (NASA), and VERITAS (NASA) missions. We hope that the upcoming data as well as advances in theoretical/numerical models will provide new insights into Venus' formation and evolution history and a deeper understanding of its divergent evolutionary trajectory from Earth. These findings will also deepen our understanding of the diversity of terrestrial planets, both within our Solar System and beyond. 

\begin{acknowledgements}
We thank the anonymous reviewer for valuable comments.
We thank Simon Grimm and Pascal Rosenblatt for valuable discussions.
The authors acknowledge the financial support of the SNSF under grant 10.001.720. We also acknowledge access to Alps.Eiger at the Swiss National Supercomputing Centre, Switzerland under the University of Zurich's share with the project ID UZH4. This work was supported by a grant from the Swiss National Supercomputing Centre (CSCS) under project ID S1285 on Piz Daint.
\end{acknowledgements}

\bibliographystyle{aa}
\bibliography{venus_paper}

\begin{thebibliography}{94}
\expandafter\ifx\csname natexlab\endcsname\relax\def\natexlab#1{#1}\fi

\bibitem[{Abbott \& Isley(2002)}]{abbott2002}
Abbott, D.~H. \& Isley, A.~E. 2002, EPSL, 205, 53

\bibitem[{Abe(1993)}]{abe1993}
Abe, Y. 1993, Thermal Evolution and Chemical Differentiation of the Terrestrial Magma Ocean (AGU), 41--54

\bibitem[{Abe(1997)}]{abe1997}
Abe, Y. 1997, Phys. Earth Planet. Inter., 100, 27

\bibitem[{Alemi \& Stevenson(2006)}]{alemiWhyVenusHas2006}
Alemi, A. \& Stevenson, D. 2006, AAS, DPS meeting \#38

\bibitem[{Asphaug \& Reufer(2014)}]{asphaug_mercury_2014}
Asphaug, E. \& Reufer, A. 2014, Nat. Geosci., 7, 564

\bibitem[{Ballantyne {et~al.}(2023)Ballantyne, Jutzi, Golabek, Mishra, Cheng, Rozel, \& Tackley}]{ballantyneInvestigatingFeasibilityImpactinduced2023}
Ballantyne, H.~A., Jutzi, M., Golabek, G.~J., {et~al.} 2023, Icarus, 392, 115395

\bibitem[{Benz {et~al.}(1988)Benz, Slattery, \& Cameron}]{benzCollisionalStrippingMercurys1988}
Benz, W., Slattery, W.~L., \& Cameron, A. 1988, Icarus, 74, 516

\bibitem[{Borgeat \& Tackley(2022)}]{borgeat2022}
Borgeat, X. \& Tackley, P.~J. 2022, Prog. Earth Planet. Sci., 9, 38

\bibitem[{Cameron \& Ward(1976)}]{cameronOriginMoon1976}
Cameron, A. G.~W. \& Ward, W.~R. 1976, Lunar Planet. Sci. Conf. Abstracts, 7, 120

\bibitem[{Canup \& Salmon(2018)}]{canupOriginPhobosDeimos2018}
Canup, R. \& Salmon, J. 2018, Sci. Adv., 4, eaar6887

\bibitem[{Canup(2012)}]{canup_forming_2012}
Canup, R.~M. 2012, Science, 338, 1052

\bibitem[{Canup \& Asphaug(2001)}]{canupOriginMoonGiant2001}
Canup, R.~M. \& Asphaug, E. 2001, Nature, 412, 708

\bibitem[{Canup \& Ward(2001)}]{canupScalingRelationshipSatelliteForming2001}
Canup, R.~M. \& Ward, W.~R. 2001, Icarus, 150

\bibitem[{Chambers(2001)}]{chambersMakingMoreTerrestrial2001}
Chambers, J.~E. 2001, Icarus, 152, 205

\bibitem[{Charnoz {et~al.}(2010)Charnoz, Salmon, \& Crida}]{charnoz_recent_2010}
Charnoz, S., Salmon, J., \& Crida, A. 2010, Nature, 465, 752

\bibitem[{Chau {et~al.}(2018)Chau, Reinhardt, Helled, \& Stadel}]{chauFormingMercuryGiant2018}
Chau, A., Reinhardt, C., Helled, R., \& Stadel, J.~G. 2018, ApJ, 865, 35

\bibitem[{Chen \& Nimmo(2016)}]{chenTidalDissipationLunar2016}
Chen, E. M.~A. \& Nimmo, F. 2016, Icarus, 275, 132

\bibitem[{Cheng {et~al.}(2024{\natexlab{a}})Cheng, Ballantyne, Golabek, Jutzi, Rozel, \& Tackley}]{chengCombinedImpactInterior2024}
Cheng, K.~W., Ballantyne, H.~A., Golabek, G.~J., {et~al.} 2024{\natexlab{a}}, Icarus, 420, 116137

\bibitem[{Cheng {et~al.}(2024{\natexlab{b}})Cheng, Ballantyne, Golabek, Jutzi, Rozel, \& Tackley}]{cheng2024combined}
Cheng, K.~W., Ballantyne, H.~A., Golabek, G.~J., {et~al.} 2024{\natexlab{b}}, Icarus, 420, 116137

\bibitem[{Correia \& Laskar(2001)}]{correiaFourFinalRotation2001}
Correia, A. C.~M. \& Laskar, J. 2001, Nature, 411, 767

\bibitem[{Correia \& Laskar(2003)}]{correiaLongtermEvolutionSpin2003a}
Correia, A. C.~M. \& Laskar, J. 2003, Icarus, 163, 24

\bibitem[{Correia {et~al.}(2003)Correia, Laskar, \& {de Surgy}}]{correiaLongtermEvolutionSpin2003}
Correia, A. C.~M., Laskar, J., \& {de Surgy}, O.~N. 2003, Icarus, 163, 1

\bibitem[{Croft(1982)}]{Croft1982}
Croft, S.~K. 1982

\bibitem[{{\'C}uk {et~al.}(2016){\'C}uk, Hamilton, Lock, \& Stewart}]{cukTidalEvolutionMoon2016}
{\'C}uk, M., Hamilton, D.~P., Lock, S.~J., \& Stewart, S.~T. 2016, Nature, 539, 402

\bibitem[{{\'C}uk {et~al.}(2021){\'C}uk, Lock, Stewart, \& Hamilton}]{cukTidalEvolutionEarthMoon2021}
{\'C}uk, M., Lock, S.~J., Stewart, S.~T., \& Hamilton, D.~P. 2021, PSJ, 2, 147

\bibitem[{{\'C}uk \& Stewart(2012)}]{cukMakingMoonFastspinning2012}
{\'C}uk, M. \& Stewart, S.~T. 2012, Science, 338, 1047

\bibitem[{Dehnen \& Aly(2012)}]{dehnenImprovingConvergenceSmoothed2012}
Dehnen, W. \& Aly, H. 2012, MNRAS, 425, 1068

\bibitem[{Denton {et~al.}(2025)Denton, Asphaug, Emsenhuber, \& Melikyan}]{denton2025}
Denton, C.~A., Asphaug, E., Emsenhuber, A., \& Melikyan, R. 2025, Nat. Geosci., 18, 37

\bibitem[{Donahue {et~al.}(1982)Donahue, Hoffman, Hodges, \& Watson}]{donahueVenusWasWet1982}
Donahue, T.~M., Hoffman, J.~H., Hodges, R.~R., \& Watson, A.~J. 1982, Science, 216, 630

\bibitem[{{Dou} {et~al.}(2024){Dou}, {Carter}, {Lock}, \& {Leinhardt}}]{Dou2024}
{Dou}, J., {Carter}, P.~J., {Lock}, S., \& {Leinhardt}, Z.~M. 2024, \mnras, 534, 758

\bibitem[{Emsenhuber {et~al.}(2023)Emsenhuber, Asphaug, Cambioni, Gabriel, \& Schwartz}]{emsenhuberCollisionChainsTerrestrial2023}
Emsenhuber, A., Asphaug, E., Cambioni, S., Gabriel, T. S.~J., \& Schwartz, S.~R. 2023, PSJ

\bibitem[{Gillmann {et~al.}(2009)Gillmann, Chassefi{\`e}re, \& Lognonn{\'e}}]{gillmannConsistentPictureEarly2009}
Gillmann, C., Chassefi{\`e}re, E., \& Lognonn{\'e}, P. 2009, EPSL, 286, 503

\bibitem[{Gillmann {et~al.}(2016)Gillmann, Golabek, \& Tackley}]{gillmannEffectSingleLarge2016}
Gillmann, C., Golabek, G.~J., \& Tackley, P.~J. 2016, Icarus, 268, 295

\bibitem[{Gillmann {et~al.}(2022)Gillmann, Way, Avice, Breuer, Golabek, Honing, {Krissansen-Totton}, Lammer, O'Rourke, Persson, Plesa, Salvador, Scherf, \& Zolotov}]{gillmannLongtermEvolutionAtmosphere2022}
Gillmann, C., Way, M.~J., Avice, G., {et~al.} 2022, Space Sci. Rev., 218, 56

\bibitem[{Hamano {et~al.}(2013)Hamano, Abe, \& Genda}]{hamanoEmergenceTwoTypes2013}
Hamano, K., Abe, Y., \& Genda, H. 2013, Nature, 497, 607

\bibitem[{Hartmann \& Davis(1975)}]{hartmannSatellitesizedPlanetesimalsLunar1975}
Hartmann, W.~K. \& Davis, D.~R. 1975, Icarus, 24

\bibitem[{Herzberg {et~al.}(2000)Herzberg, Raterron, \& Zhang}]{herzberg2000new}
Herzberg, C., Raterron, P., \& Zhang, J. 2000, Geochem. Geophys. Geosyst., 1

\bibitem[{Hosono {et~al.}(2017)Hosono, Iwasawa, Tanikawa, Nitadori, Muranushi, \& Makino}]{hosono_unconvergence_2017}
Hosono, N., Iwasawa, M., Tanikawa, A., {et~al.} 2017, PASJ, 69, 26

\bibitem[{Hyodo {et~al.}(2017)Hyodo, Genda, Charnoz, \& Rosenblatt}]{hyodoImpactOriginPhobos2017}
Hyodo, R., Genda, H., Charnoz, S., \& Rosenblatt, P. 2017, ApJ, 845, 125

\bibitem[{Hyodo {et~al.}(2015)Hyodo, Ohtsuki, \& Takeda}]{hyodo_formation_2015}
Hyodo, R., Ohtsuki, K., \& Takeda, T. 2015, ApJ, 799, 40

\bibitem[{Jones {et~al.}(2002)Jones, Price, Price, DeCarli, \& Clegg}]{Jones2002}
Jones, A.~P., Price, G.~D., Price, N.~J., DeCarli, P.~S., \& Clegg, R.~A. 2002, EPSL, 202, 551

\bibitem[{Kaula(1979)}]{Kaula1979}
Kaula, W.~M. 1979, J. Geophys. Res. Solid Earth, 84, 999

\bibitem[{Kegerreis {et~al.}(2018)Kegerreis, Teodoro, Eke, Massey, Catling, Fryer, Korycansky, Warren, \& Zahnle}]{kegerreis_consequences_2018}
Kegerreis, J.~A., Teodoro, L. F.~A., Eke, V.~R., {et~al.} 2018, ApJ, 861, 52

\bibitem[{Lay {et~al.}(2008)Lay, Hernlund, \& Buffett}]{lay2008cmb}
Lay, T., Hernlund, J., \& Buffett, B.~A. 2008, Nat. Geosci., 1, 25

\bibitem[{Lissauer \& Kary(1991)}]{lissauerOriginSystematicComponent1991}
Lissauer, J.~J. \& Kary, D.~M. 1991, Icarus, 94, 126

\bibitem[{Lock \& Stewart(2017)}]{lock_structure_2017}
Lock, S.~J. \& Stewart, S.~T. 2017, J. Geophys. Res. Planets, 122, 950

\bibitem[{Louren{\c{c}}o {et~al.}(2020)Louren{\c{c}}o, Rozel, Ballmer, \& Tackley}]{lourencco2020plutonic}
Louren{\c{c}}o, D.~L., Rozel, A.~B., Ballmer, M.~D., \& Tackley, P.~J. 2020, Geochem. Geophys. Geosyst., 21, e2019GC008756

\bibitem[{Marchi {et~al.}(2023)Marchi, Rufu, \& Korenaga}]{marchiLonglivedVolcanicResurfacing2023}
Marchi, S., Rufu, R., \& Korenaga, J. 2023, Nat. Astron.

\bibitem[{Margot {et~al.}(2021)Margot, Campbell, Giorgini, Jao, Snedeker, Ghigo, \& Bonsall}]{margotSpinStateMoment2021}
Margot, J.-L., Campbell, D.~B., Giorgini, J.~D., {et~al.} 2021, Nat. Astron., 5, 676

\bibitem[{Maruyama {et~al.}(2018)Maruyama, Santosh, \& Azuma}]{maruyama2018}
Maruyama, S., Santosh, M., \& Azuma, S. 2018, Geosci. Front., 9, 1033

\bibitem[{McDonough \& Sun(1995)}]{mcdonough1995composition}
McDonough, W.~F. \& Sun, S.-S. 1995, Chem. Geol., 120, 223

\bibitem[{Meier \& Reinhardt(2021)}]{meierEOSlib2021}
Meier, T. \& Reinhardt, C. 2021, {{EOSlib}}

\bibitem[{{Meier} {et~al.}(2025){Meier}, {Reinhardt}, {Shibata}, {M{\"u}ller}, {Stadel}, \& {Helled}}]{Meier2025}
{Meier}, T., {Reinhardt}, C., {Shibata}, S., {et~al.} 2025, ApJ, 988, 7

\bibitem[{Meier {et~al.}(2021)Meier, Reinhardt, \& Stadel}]{meierEOSResolutionConspiracy2021}
Meier, T., Reinhardt, C., \& Stadel, J. 2021, MNRAS, 505, 1806

\bibitem[{Meier {et~al.}(2024)Meier, Reinhardt, Timpe, Stadel, \& Moore}]{meierSystematicSurveyMoonforming2025}
Meier, T., Reinhardt, C., Timpe, M., Stadel, J., \& Moore, B. 2024, ApJ, 978, 11

\bibitem[{Melosh(1989)}]{melosh_impact_1989}
Melosh, H.~J. 1989, Impact cratering: {A} geologic process

\bibitem[{Melosh(2007)}]{meloshHydrocodeEquationState2007}
Melosh, H.~J. 2007, Meteorit. Planet. Sci., 42, 2079

\bibitem[{Monaghan(1992)}]{monaghanSmoothedParticleHydrodynamics1992}
Monaghan, J.~J. 1992, Annu. Rev. Astron. Astrophys., 30, 543

\bibitem[{Monteux {et~al.}(2016)Monteux, Andrault, \& Samuel}]{monteux2017magmaocean}
Monteux, J., Andrault, D., \& Samuel, H. 2016, EPSL

\bibitem[{Monteux {et~al.}(2007)Monteux, Coltice, Dubuffet, \& Ricard}]{monteux2007}
Monteux, J., Coltice, N., Dubuffet, F., \& Ricard, Y. 2007, Geophys. Res. Lett., 34

\bibitem[{Morbidelli {et~al.}(2025)Morbidelli, Kleine, \& Nimmo}]{morbidelliDidTerrestrialPlanets2024}
Morbidelli, A., Kleine, T., \& Nimmo, F. 2025, EPSL, 650, 119120

\bibitem[{Musseau {et~al.}(2024)Musseau, Tobie, Dumoulin, Gillmann, Revol, \& Bolmont}]{musseauViscosityVenusMantle2024}
Musseau, Y., Tobie, G., Dumoulin, C., {et~al.} 2024, Icarus, 422, 116245

\bibitem[{Okeefe \& Ahrens(1975)}]{okeefe1975}
Okeefe, J.~D. \& Ahrens, T.~J. 1975, in Lunar Planet. Sci. Conf. Abstracts, 6th, Vol. 3, Vol.~6, 2831--2844

\bibitem[{O’Neill {et~al.}(2017)O’Neill, Marchi, Zhang, \& Bottke}]{oneill2017}
O’Neill, C., Marchi, S., Zhang, S., \& Bottke, W. 2017, Nat. Geosci., 10, 793

\bibitem[{Potter {et~al.}(2017)Potter, Stadel, \& Teyssier}]{potterPKDGRAV3TrillionParticle2017}
Potter, D., Stadel, J., \& Teyssier, R. 2017, Comput. Astrophys. Cosm., 4, 2

\bibitem[{Price(2012)}]{priceSmoothedParticleHydrodynamics2012}
Price, D.~J. 2012, J. Comput. Phys., 231, 759

\bibitem[{Quintana {et~al.}(2016)Quintana, Barclay, Borucki, Rowe, \& Chambers}]{quintanaFrequencyGiantImpacts2016}
Quintana, E.~V., Barclay, T., Borucki, W.~J., Rowe, J.~F., \& Chambers, J.~E. 2016, ApJ, 821, 126

\bibitem[{Raymond {et~al.}(2006)Raymond, Quinn, \& Lunine}]{raymondHighresolutionSimulationsFinal2006}
Raymond, S.~N., Quinn, T., \& Lunine, J.~I. 2006, Icarus, 183, 265

\bibitem[{Reese {et~al.}(2004)Reese, Solomatov, Baumgardner, Stegman, \& Vezolainen}]{reese2004}
Reese, C., Solomatov, V.~S., Baumgardner, J., Stegman, D.~R., \& Vezolainen, A. 2004, J. Geophys. Res. Planets, 109

\bibitem[{Reinhardt(2020)}]{reinhardtSimulatingGiantImpacts2020}
Reinhardt, C. 2020, PhD thesis, University of Zurich

\bibitem[{Reinhardt {et~al.}(2020)Reinhardt, Chau, Stadel, \& Helled}]{reinhardtBifurcationHistoryUranus2020}
Reinhardt, C., Chau, A., Stadel, J., \& Helled, R. 2020, MNRAS, 492, 5336

\bibitem[{Reinhardt {et~al.}(2022)Reinhardt, Meier, Stadel, Otegi, \& Helled}]{reinhardt_forming_2022}
Reinhardt, C., Meier, T., Stadel, J.~G., Otegi, J.~F., \& Helled, R. 2022, MNRAS, 517, 3132

\bibitem[{Reinhardt \& Stadel(2017)}]{reinhardtNumericalAspectsGiant2017}
Reinhardt, C. \& Stadel, J. 2017, MNRAS, 467, 4252

\bibitem[{Revol {et~al.}(2023)Revol, Bolmont, Tobie, Dumoulin, Musseau, Mathis, Strugarek, \& Brun}]{revolSpinEvolutionVenuslike2023}
Revol, A., Bolmont, {\'E}., Tobie, G., {et~al.} 2023, A\&A, 674, A227

\bibitem[{Roberts \& Barnouin(2012)}]{Roberts2012}
Roberts, J.~H. \& Barnouin, O.~S. 2012, J. Geophys. Res. Planets, 117

\bibitem[{Rosenblatt {et~al.}(2016)Rosenblatt, Charnoz, Dunseath, {Terao-Dunseath}, Trinh, Hyodo, Genda, \& Toupin}]{rosenblattAccretionPhobosDeimos2016}
Rosenblatt, P., Charnoz, S., Dunseath, K.~M., {et~al.} 2016, Nat. Geosci., 9, 581

\bibitem[{Ruedas \& Breuer(2019)}]{ruedas2019}
Ruedas, T. \& Breuer, D. 2019, Phys. Earth Planet. Inter., 287, 76

\bibitem[{Rufu \& Canup(2020)}]{rufuTidalEvolutionEvection2020}
Rufu, R. \& Canup, R.~M. 2020, J. Geophys. Res. Planets, 125, e2019JE006312

\bibitem[{{Ruiz-Bonilla} {et~al.}(2022){Ruiz-Bonilla}, Borrow, Eke, Kegerreis, Massey, Sandnes, \& Teodoro}]{ruiz-bonillaDealingDensityDiscontinuities2022}
{Ruiz-Bonilla}, S., Borrow, J., Eke, V.~R., {et~al.} 2022, MNRAS, 512, 4660

\bibitem[{Safronov(1978)}]{Safronov1978}
Safronov, V. 1978, Icarus, 33, 3

\bibitem[{Springel \& Hernquist(2002)}]{springelCosmologicalSmoothedParticle2002}
Springel, V. \& Hernquist, L. 2002, MNRAS, 333, 649

\bibitem[{Stadel(2001)}]{stadelCosmologicalNbodySimulations2001}
Stadel, J.~G. 2001, PhD thesis, University of Washington

\bibitem[{Stewart(2020)}]{stewartEquationStateModel2020}
Stewart, S.~T. 2020, Zenodo

\bibitem[{Stewart {et~al.}(2019)Stewart, Davies, Duncan, Lock, Root, Townsend, Kraus, Caracas, \& Jacobsen}]{stewartEquationStateModel2019}
Stewart, S.~T., Davies, E.~J., Duncan, M.~S., {et~al.} 2019, Zenodo

\bibitem[{Tackley(2008)}]{tackley2008modelling}
Tackley, P.~J. 2008, Phys. Earth Planet. Inter., 171, 7

\bibitem[{Thompson \& Lauson(1972)}]{thompsonImprovementsChartRadiationhydrodynamic1972}
Thompson, S.~L. \& Lauson, H. 1972

\bibitem[{Thompson {et~al.}(2019)Thompson, Lauson, Melosh, Collins, \& Stewart}]{thompsonMANEOS2019}
Thompson, S.~L., Lauson, H., Melosh, H.~J., Collins, G., \& Stewart, S.~T. 2019, Zenodo

\bibitem[{Tian {et~al.}(2023)Tian, Tackley, \& Louren{\c{c}}o}]{tian2023tectonics}
Tian, J., Tackley, P.~J., \& Louren{\c{c}}o, D.~L. 2023, Icarus, 399, 115539

\bibitem[{Timpe {et~al.}(2023)Timpe, Reinhardt, Meier, Stadel, \& Moore}]{timpeSystematicSurveyMoonforming2023}
Timpe, M., Reinhardt, C., Meier, T., Stadel, J., \& Moore, B. 2023, APJ, 959, 38

\bibitem[{Watters {et~al.}(2009)Watters, Zuber, \& Hager}]{watters2009}
Watters, W.~A., Zuber, M., \& Hager, B. 2009, J. Geophys. Res. Planets, 114

\bibitem[{Way {et~al.}(2020)Way, Anthony, \& Del~Genio}]{wayVenusianHabitableClimate2020}
Way, M.~J., Anthony, \& Del~Genio, D. 2020, J. Geophys. Res. Planets, 125

\bibitem[{Wilhelms \& Squyres(1984)}]{wilhelmsMartianHemisphericDichotomy1984}
Wilhelms, D.~E. \& Squyres, S.~W. 1984, Nature, 309, 138

\bibitem[{Woo {et~al.}(2022)Woo, Reinhardt, Cilibrasi, Chau, Helled, \& Stadel}]{woo_did_2022}
Woo, J. M.~Y., Reinhardt, C., Cilibrasi, M., {et~al.} 2022, Icarus, 375, 114842

\bibitem[{Zerr {et~al.}(1998)Zerr, Diegeler, \& Boehler}]{zerr1998solidus}
Zerr, A., Diegeler, A., \& Boehler, R. 1998, Science, 281, 243

\end{thebibliography}

\begin{appendix}
\onecolumn
\section{Collisions table}

In total, we run 81 impact simulations that are displayed in the following table.
We find 64 merger, 7 graze-and-merger, and 10 hit-and-run collisions.
Simulations highlighted in green have post-impact rotation periods of \SI{48}{\hour} or higher and are consistent with Venus' present-day rotation period.

\begin{table}[ht]
\centering
\caption{Summary of all collisions.}
    \resizebox{440pt}{!}{
    \begin{tabular}{lrrrrlrlrrrrr} 
    \\
    {ID} & {$b$} & {$v_{imp} \ [\SI{}{\kilo\meter\per\second}]$} & {Impactor Mass [\SI{}{\ME}]} &              \multicolumn{3}{c}{Pre-impact Venus}          & {Outcome Scenario} & \multicolumn{3}{c}{Post-impact Venus}     & {Disc} & {$R_{syn}$}\\
    \\ 
         &  &                           &                            &  Mass [\SI{}{\ME}] & {Target Model} & {Period [\SI{}{\hour}]} &                    & {M [\SI{}{\ME}]} & {Period [\SI{}{\hour}]} & {angular momentum [$L_{em}$]} &  {M [\SI{}{\MV}]} & {M [\SI{}{\MV}]} \\ 
     \rowcolor{green!15}
     1  &   0.0    & 10.0 & 0.1  & 0.715 & Cold & 0   & {merger}            & 0.812 & 20000     & 0.00      & 0 & 0\\
     2  &   0.3    & 10.0 & 0.1  & 0.715 & Cold & 0   & {merger}            & 0.809 & 9.4       & 0.35      & 0 & 0\\
     3  &   0.7    & 10.0 & 0.1  & 0.750 & Cold & 0   & {graze-and-merge}  & 0.824 & 5.8       & 0.60      & \num{2.4e-3} & \num{2.1e-3}\\
     \rowcolor{green!15}
     4  &   0.0    & 15.0 & 0.1  & 0.715 & Cold & 0   & {merger}            & 0.800 & 5700    & 0.00      & 0 & 0 \\
     5  &   0.3    & 15.0 & 0.1  & 0.750 & Cold & 0   & {merger}            & 0.823 & 9.6       & 0.42      & \num{1.8e-3} & 0\\
     6  &   0.7    & 15.0 & 0.1  & 0.815 & Cold & 0   & {hit-and-run}       & 0.828 & 13.6      & 0.24      & \num{0.3e-3} & 0\\  
        
     7  &   0.0    & 10.0 & 0.1  & 0.715 & Cold & 12  & {merger}            & 0.813 & 16.2      & 0.20      & 0 & 0\\
     8  &   0.3    & 10.0 & 0.1  & 0.715 & Cold & 12  & {merger}            & 0.809 & 6.1       & 0.56      & 0 & 0\\
     9  &   0.7    & 10.0 & 0.1  & 0.750 & Cold & 12  & {graze-and-merge}  & 0.826 & 4.5       & 0.83      & \num{4.5e-3} & \num{4.2e-3}\\
     10  &   -0.3   & 10.0 & 0.1  & 0.715 & Cold & 12  & {merger}            & 0.808 & -21.9     & -0.15     & 0 & 0\\
     11  &   -0.7   & 10.0 & 0.1  & 0.750 & Cold & 12  & {graze-and-merge}  & 0.823 & -8.9      & -0.40     & \num{1.9e-3} & \num{1.0e-3}\\
     12   &   0.0    & 15.0 & 0.1  & 0.715 & Cold & 12  & {merger}            & 0.800 & 19.8      & 0.19      & 0 & 0\\
     13  &   0.3    & 15.0 & 0.1  & 0.750 & Cold & 12  & {merger}            & 0.823 & 6.7       & 0.63      & \num{3.3e-3} & \num{1.8e-3}\\
     14  &   0.7    & 15.0 & 0.1  & 0.815 & Cold & 12  & {hit-and-run}       & 0.825 & 7.2       & 0.45      & \num{0.3e-3} & \num{0.2e-3}\\ 
     15  &   -0.3   & 15.0 & 0.1  & 0.750 & Cold & 12  & {merger}            & 0.824 & -18.9     & -0.22     & \num{0.4e-3} & 0\\
     \rowcolor{green!15}
     16  &   -0.7   & 15.0 & 0.1  & 0.815 & Cold & 12  & {hit-and-run}       & 0.830 & -111.5    & -0.03     & \num{0.2e-3} & 0\\ 

     \rowcolor{green!15}
     17  &   -0.7   & 10.0 & 0.1  & 0.750 & Cold & 6   & {hit-and-run}            & 0.807 & -69.1     & 0.05      & \num{1.2e-3} & 0\\ 

     18  &   -0.3   & 10.0 & 0.1  & 0.715 & Cold & 24  & {merger}            & 0.808 & -13.1     & 0.25      & \num{0.2e-3} & 0\\
     19  &   -0.7   & 10.0 & 0.1  & 0.750 & Cold & 24  & {graze-and-merge}  & 0.823 & -7.0      & 0.49      & \num{3.2e-3} & \num{1.7e-3}\\
     20  &   -0.3   & 15.0 & 0.1  & 0.750 & Cold & 24  & {merger}            & 0.824 & -12.5     & 0.32      & \num{0.9e-3} & 0\\
     21  &   -0.7   & 15.0 & 0.1  & 0.815 & Cold & 24  & {hit-and-run}       & 0.828 & -24.4     & 0.13      & \num{0.3e-3} & 0.00\\ 

     22  &   -0.3   & 10.0 & 0.1  & 0.715 & Cold & 2.5 & {merger}            & 0.796 & 3.8       & 1.13      & \num{9.1e-3} & \num{9.1e-3}\\
     23  &    -0.7   & 10.0 & 0.1  & 0.750 & Cold & 2.5 & {hit-and-run}       & 0.798 & 3.7       & 1.08      & \num{4.7e-3} & \num{4.7e-3}\\
     24  &    -0.3   & 15.0 & 0.1  & 0.750 & Cold & 2.5 & {merger}            & 0.809 & 4.6       & 1.09      & \num{9.5e-3} & \num{9.5e-3}\\
     25  &    -0.7   & 15.0 & 0.1  & 0.815 & Cold & 2.5 & {hit-and-run}       & 0.834 & 3.3       & 1.37      & \num{7.7e-3} & \num{7.7e-3}\\ 

     \rowcolor{green!15}
     26  &   0.0    & 10.0 & 0.1  & 0.715 & Hot  & 0   & {merger}            & 0.812 & 37400     & 0.00      & 0  & 0\\
     27  &   0.3    & 10.0 & 0.1  & 0.715 & Hot  & 0   & {merger}            & 0.809 &  9.5      & 0.35      & 0  & 0  \\
     28  &   0.7    & 10.0 & 0.1  & 0.750 & Hot  & 0   & {graze-and-merge}  & 0.823 & 5.9       & 0.60      & \num{2.7e-3}  & \num{2.3e-3}\\
     \rowcolor{green!15}
     29  &   0.0    & 15.0 & 0.1  & 0.715 & Hot  & 0   & {merger}            & 0.800 & 5600      & 0.00      & 0  & 0\\
     30  &   0.3    & 15.0 & 0.1  & 0.750 & Hot  & 0   & {merger}            & 0.823 & 9.7       & 0.42      & \num{1.8e-3}  & \num{0.5e-3}\\
     31  &   0.7    & 15.0 & 0.1  & 0.815 & Hot  & 0   & {hit-and-run}       & 0.827 & 13.5      & 0.24      & \num{0.3e-3}  & 0\\ 

     32  &   0.0    & 10.0 & 0.1  & 0.715 & Hot  & 12  & {merger}            & 0.813  & 16.3     & 0.20      &  0 & 0\\
     33  &   0.3    & 10.0 & 0.1  & 0.715 & Hot  & 12  & {merger}            & 0.809  & 6.1      & 0.56      &  0 & 0\\
     34  &   0.7    & 10.0 & 0.1  & 0.750 & Hot  & 12  & {graze-and-merge}  & 0.825  & 4.5      & 0.82      &  \num{5.1e-3} & \num{5.0e-3}\\
     35  &   -0.3   & 10.0 & 0.1  & 0.715 & Hot  & 12  & {merger}            & 0.808  & -22.0    & -0.15     &  \num{0.2e-3} & 0\\
     36  &   -0.7   & 10.0 & 0.1  & 0.750 & Hot  & 12  & {graze-and-merge}  & 0.823  & -8.9     & -0.40     &  \num{2.0e-3} & \num{1.2e-3}\\
     37  &   0.0    & 15.0 & 0.1  & 0.715 & Hot  & 12  & {merger}            & 0.800  & 20.1     & 0.19      &  \num{0.2e-3} & 0\\
     38  &   0.3    & 15.0 & 0.1  & 0.750 & Hot  & 12  & {merger}            & 0.822  & 6.7      & 0.63      &  \num{3.6e-3} & \num{2.0e-3}\\
     39  &   0.7    & 15.0 & 0.1  & 0.815 & Hot  & 12  & {hit-and-run}       & 0.823  & 7.3      & 0.44      &  \num{0.3e-3} & \num{0.2e-3}\\ 
     40  &   -0.3   & 15.0 & 0.1  & 0.750 & Hot  & 12  & {merger}            & 0.824  & -19.0    & -0.22     &  \num{0.6e-3} & 0\\
     \rowcolor{green!15}
     41  &   -0.7   & 15.0 & 0.1  & 0.815 & Hot  & 12  & {hit-and-run}       & 0.830  & -106.4   & -0.03     &  \num{0.3e-3} & 0\\ 

     \rowcolor{green!15}
     42  &   0.0    & 10.0 & 0.01 & 0.805 & Cold & 0   &  {merger}   & 0.815  & 2900   & 0.00      &  0 &  0\\
     \rowcolor{green!15}
     43  &   0.3    & 10.0 & 0.01 & 0.805 & Cold & 0   &  {merger}   & 0.815  & 83.8     & 0.04      &  0 &  0\\
     44  &   0.7    & 10.0 & 0.01 & 0.805 & Cold & 0   &  {merger}   & 0.814  & 42.1     & 0.07      &  0 \\
     \rowcolor{green!15}
     45  &   0.0    & 15.0 & 0.01 & 0.805 & Cold & 0   &  {merger}   & 0.815  & -1900    & 0.00      &  0 &  0\\
     \rowcolor{green!15}
     46  &   0.3    & 15.0 & 0.01 & 0.805 & Cold & 0   &  {merger}   & 0.814  & 58.7     & 0.05      &  0 &  0\\
     47  &   0.7    & 15.0 & 0.01 & 0.815 & Cold & 0   &  {merger}   & 0.821  & 38.2     & 0.08      &  0 &  0\\ 

     48  &   0.0    & 10.0 & 0.01 & 0.805 & Cold & 12  &  {merger}   & 0.815 & 12.6      & 0.24      &  0 &  0\\
     49  &   0.3    & 10.0 & 0.01 & 0.805 & Cold & 12  &  {merger}   & 0.815 & 10.9      & 0.27      &  0 &  0\\
     50  &   0.7    & 10.0 & 0.01 & 0.805 & Cold & 12  &  {merger}   & 0.814 & 9.7       & 0.31      &  0 &  0\\
     51  &   -0.3   & 10.0 & 0.01 & 0.805 & Cold & 12  &  {merger}   & 0.815 & 14.8      & 0.20      &  0 &  0\\
     52  &   -0.7   & 10.0 & 0.01 & 0.805 & Cold & 12  &  {merger}   & 0.813 & 17.5      & 0.17      &  0 &  0\\
     53  &   0.0    & 15.0 & 0.01 & 0.805 & Cold & 12  &  {merger}   & 0.815 & 12.7      & 0.24      &  0 &  0\\
     54  &   0.3    & 15.0 & 0.01 & 0.805 & Cold & 12  &  {merger}   & 0.814 & 10.4      & 0.29      &  0 &  0\\
     55  &   0.7    & 15.0 & 0.01 & 0.805 & Cold & 12  &  {merger}   & 0.811 & 9.5       & 0.32      &  0 &  0\\ 
     56  &   -0.3   & 15.0 & 0.01 & 0.805 & Cold & 12  &  {merger}   & 0.814 & 16.2      & 0.18      &  0 &  0\\
     57  &   -0.7   & 15.0 & 0.01 & 0.805 & Cold & 12  &  {merger}   & 0.811 & 18.3      & 0.16      &  0 &  0\\ 

     58  &   -0.3   & 10.0 & 0.01 & 0.805 & Cold & 24  &  {merger}   & 0.815 & 36.2      & 0.08      &  0 &  0\\
     \rowcolor{green!15}
     59  &   -0.7   & 10.0 & 0.01 & 0.805 & Cold & 24  &  {merger}   & 0.813 & 60.0      & 0.05      &  00 &  0\\
     60  &   -0.3   & 15.0 & 0.01 & 0.805 & Cold & 24  &  {merger}   & 0.814 & 45.2      & 0.07      &  0 &  0\\
     \rowcolor{green!15}
     61  &   -0.7   & 15.0 & 0.01 & 0.805 & Cold & 24  &  {merger}   & 0.810 & 67.8      & 0.04      &  0 &  0\\ 

     62  &   -0.3   & 10.0 & 0.01 & 0.805 & Cold & 2.5  &  {merger}   & 0.814 & 2.6      & 1.75      &  \num{2.6e-3} &  \num{2.6e-3}\\
     63  &   -0.7   & 10.0 & 0.01 & 0.805 & Cold & 2.5  &  {merger}   & 0.812 & 2.6      & 1.72      &  \num{2.7e-3} &  \num{2.7e-3}\\
     64  &   -0.3   & 15.0 & 0.01 & 0.805 & Cold & 2.5  &  {merger}   & 0.813 & 2.7      & 1.72      &  \num{6.0e-3} &  \num{6.0e-3}\\
     65  &   -0.7   & 15.0 & 0.01 & 0.805 & Cold & 2.5  &  {merger}   & 0.809 & 2.7      & 1.70      &  \num{3.8e-3} &  \num{3.8e-3}\\ 

     \rowcolor{green!15}
     66  &   0.0    & 10.0 & 0.01 & 0.805 & Hot  & 0   &  {merger}   & 0.815 & -2900   & 0.00      &  0 &  0\\
     \rowcolor{green!15}
     67  &   0.3    & 10.0 & 0.01 & 0.805 & Hot  & 0   &  {merger}   & 0.815 & 83.6      & 0.04      &  0 &  0\\
     68  &   0.7    & 10.0 & 0.01 & 0.805 & Hot  & 0   &  {merger}   & 0.814 & 42.2      & 0.07      &  0 &  0\\
     \rowcolor{green!15}
     69  &   0.0    & 15.0 & 0.01 & 0.805 & Hot  & 0   &  {merger}   & 0.815 & -1900     & 0.00      &  0 &  0\\
     \rowcolor{green!15}
     70  &   0.3    & 15.0 & 0.01 & 0.805 & Hot  & 0   &  {merger}   & 0.814 & 58.9      & 0.25      &  0 &  0\\
     71  &   0.7    & 15.0 & 0.01 & 0.815 & Hot  & 0   &  {merger}   & 0.821 & 37.9      & 0.08      &  0 &  0\\ 

     72  &   0.0    & 10.0 & 0.01 & 0.805 & Hot  & 12  &  {merger}   & 0.815 & 12.7      & 0.24      &  0 &  0\\
     73  &   0.3    & 10.0 & 0.01 & 0.805 & Hot  & 12  &  {merger}   & 0.815 & 11.0      & 0.27      &  0 &  0\\
     74  &   0.7    & 10.0 & 0.01 & 0.805 & Hot  & 12  &  {merger}   & 0.814 & 9.7       & 0.31      &  0 &  0\\
     75  &   -0.3   & 10.0 & 0.01 & 0.805 & Hot  & 12  &  {merger}   & 0.815 & 14.9      & 0.20      &  0 &  0\\
     76  &   -0.7   & 10.0 & 0.01 & 0.805 & Hot  & 12  &  {merger}   & 0.813 & 17.7      & 0.17      &  0 &  0\\
     77  &   0.0    & 15.0 & 0.01 & 0.805 & Hot  & 12  &  {merger}   & 0.815 & 12.8      & 0.24      &  0 &  0\\
     78  &   0.3    & 15.0 & 0.01 & 0.805 & Hot  & 12  &  {merger}   & 0.814 & 10.5      & 0.29      &  0 &  0\\
     79  &   0.7    & 15.0 & 0.01 & 0.805 & Hot  & 12  &  {merger}   & 0.811 & 9.5       & 0.32      &  0 &  0\\
     80  &   -0.3   & 15.0 & 0.01 & 0.805 & Hot  & 12  &  {merger}   & 0.814 & 16.3      & 0.18      &  0 &  0\\
     81  &   -0.7   & 15.0 & 0.01 & 0.805 & Hot  & 12  &  {merger}   & 0.811 & 18.4      & 0.16      &  0 &  0\\ 

    \end{tabular}
    }
    \tablefoot{The first column gives the ID of the collisions, the next six columns describe the impact conditions and the final six columns their outcomes.}
    \label{tab:collisions}
\end{table}
\newpage
\section{Post-impact angular momentum plot}

\begin{figure}
    \resizebox{0.85\hsize}{!}{%
        \includegraphics{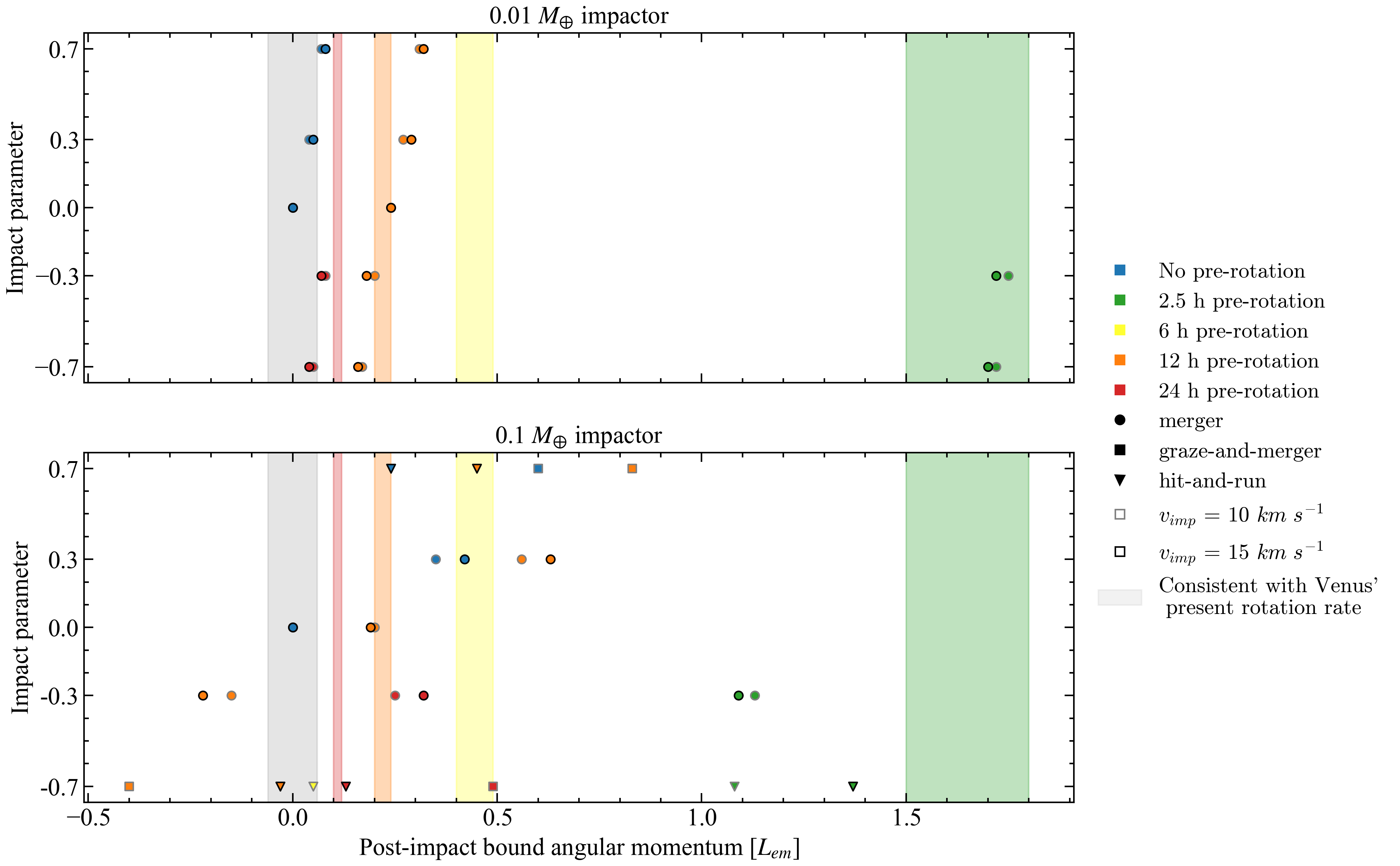}
    }
    \caption{Same as Fig. \ref{fig:periods} but using the total post-impact bound angular momentum (planet + disc) instead of rotation period.}
    \label{fig:AM_plot}
\end{figure}

After each collision we determined the post-impact rotation period and angular momentum of Venus to check compatibility. As discussed in Sec. \ref{sec:rotationrates}, the rotation period directly after a collision is expected to slightly change due to cooling and contraction. To verify that our qualitative results remain we also checked compatibility with means of the bound angular momentum (planet and disc) (see Fig.~\ref{fig:AM_plot}). The shaded region corresponds to the angular momentum of a present-day Venus with a rotation period of \SI{48}{\hour}. We find the same collisions to be consistent with Venus' present-day rotation rate.

\section{1D post-impact temperature profiles}
\begin{figure}
    \resizebox{0.9\hsize}{!}{%
        \includegraphics{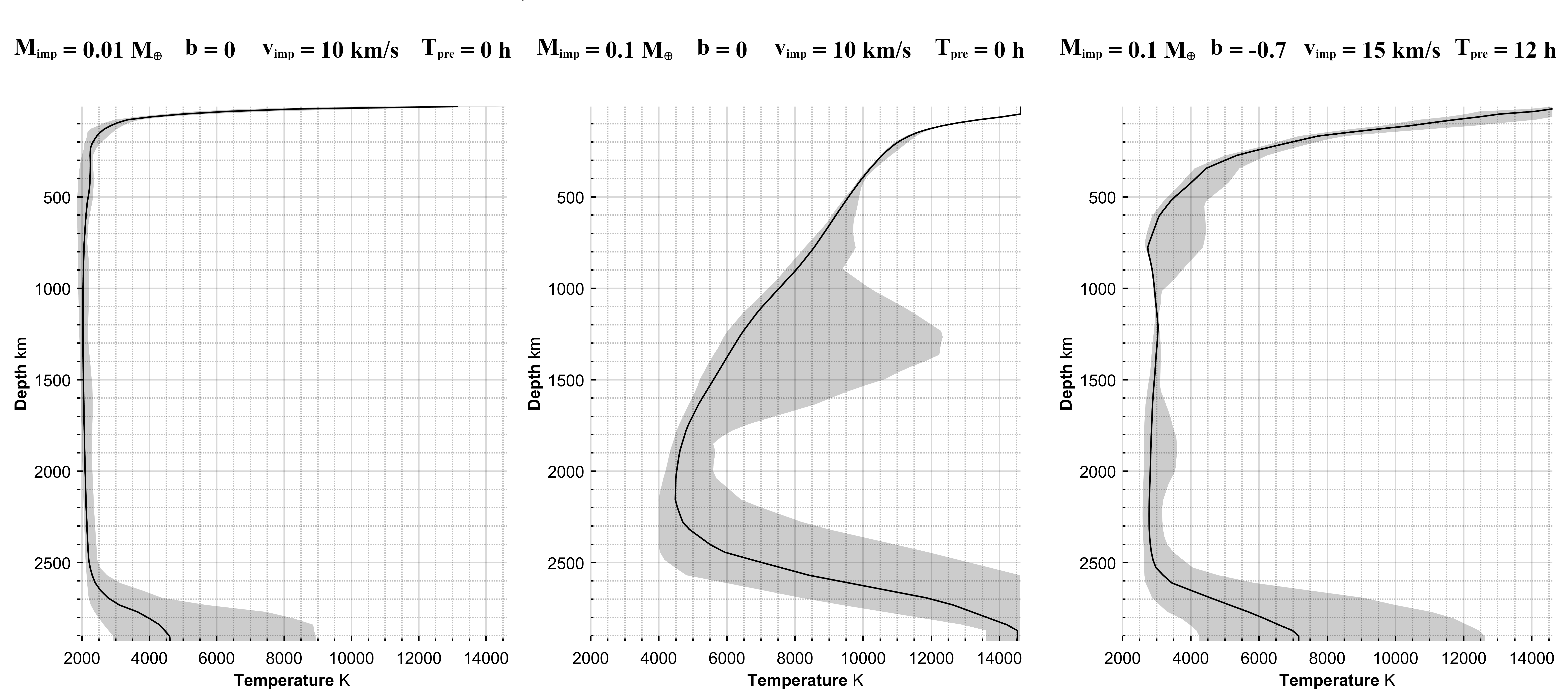}
    }
    \caption{1D mantle temperature profile for different collisions. The black lines represent the average temperature, while the grey shaded area shows the range between minimal and maximal values at each depth. The impact conditions are given at the top of each panel.}
    \label{fig:post_imp_temp_plots}
\end{figure}

Fig.~\ref{fig:post_imp_temp_plots} shows the 1D temperature profiles of three collisions that are consistent with Venus' present-day rotation rate and lack of a moon. 
It illustrates that consistent collisions on Venus are very diverse and lead to very different temperature profiles which may influence the subsequent thermal evolution.

\newpage
\section{Melting profiles of the early post-impact evolution}

\begin{figure}
    \resizebox{\hsize}{!}{%
        \includegraphics{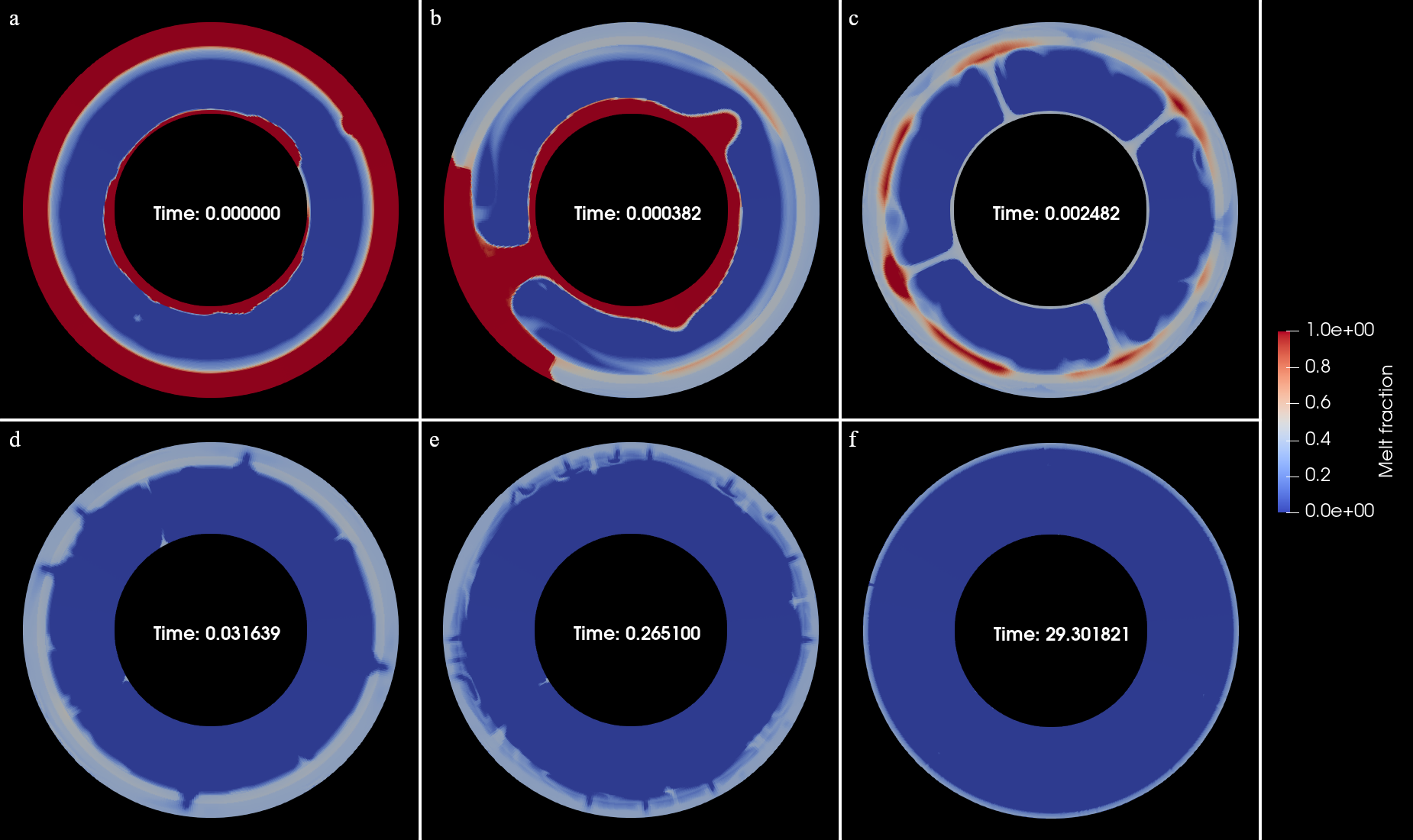}
    }
    \caption{Melt fraction field for a \texttt{StagYY} post-impact simulation featuring case c displayed on Fig.~\ref{fig:melting_profiles}. Time is in Myr. The figure illustrates the next 30 Myr of the model's evolution through the cooling stages of the planet. a) post-impact initial conditions. b) At 382 years, the surface has reached the rheological transition and the deep buoyant melt rises and freezes (end of stage i). c) At 2482 years, the deep melt layer has disappeared and pockets of melt remain in the lower regions of the upper mantle. d) At 31639 years, the layer of intermediate melt fraction (35-50\%) is still visible. e) At 0.26 Myr, melt fraction in the upper mantle is fully below the rheological transition (end of stage ii). f) By 29 Myr, the surface is starting to be fully solid (end of stage iii).}
    \label{fig:early_evolution}
\end{figure}

After the rheological transition, subsequent cooling phases are not studied in detail in this work and will be investigated in a follow-up study. However, for reference, we include a brief view of phases (ii) and (iii) in Fig.~\ref{fig:early_evolution} to highlight the timescale of the process. Then, intermediate melt fractions (above 35\%) can subsist in the upper mantle for some time. All remaining initial pockets of liquid disappear by 0.2-0.5 Myr (ii).  A strictly solid surface (0\% remaining melt fraction excluding newly generated melt) is achieved over timescales on the order of 1-50 Myr (iii).

\end{appendix}
\end{document}